\nonstopmode
\documentclass[twocolumn]{aastex631}
\usepackage[lf,scale=1.16]{ebgaramond}\def\garamond{1}\usepackage{ebgaramond-maths}\usepackage{mathtools}
\usepackage{amsmath}
\usepackage{bm}
\usepackage{hyperref}

\newcounter{myequation}
\makeatletter
\@addtoreset{equation}{myequation}
\makeatother

\usepackage{graphicx}
\usepackage{xspace}
\usepackage{gensymb}

\newcommand\shorthandon{\catcode`*=\active}
\newcommand\shorthandoff{\catcode`*=12}
\shorthandon
\def*#1 {\acold{#1}\xspace}
\shorthandoff
\begin{document}
\renewcommand{\sectionautorefname}{\S\negthinspace}
\renewcommand{\figureautorefname}{Fig.}
\renewcommand{\equationautorefname}{Eq.}
\renewcommand{\tableautorefname}{Table}
\renewcommand{\subsectionautorefname}{\S\negthinspace}
\renewcommand{\subsubsectionautorefname}{\S\negthinspace}

\title{Resolved ALMA [C{\sc ii}] 158 \micron Observations at Cosmic Noon: ISM Structure and Dynamics of Starbursting QSO SDSSJ1000}

\correspondingauthor{Christopher Rooney}
\email{ctr44@cornell.edu}

\author[0000-0002-8513-2971]{Christopher Rooney}
\affiliation{Department of Astronomy, Cornell University, Ithaca, NY 14853, USA}
\affiliation{National Institute of Standards and Technology, Boulder, CO 80305, USA}

\author[0000-0002-1605-0032]{Bo Peng}
\affiliation{Max Planck Institute for Astrophysics¸ Garching, Germany}

\author[0000-0002-4444-8929]{Amit Vishwas}
\affiliation{Cornell Center for Astrophysics and Planetary Science, Cornell University, Ithaca, NY 14853, USA}

\author[0000-0003-1260-5448]{Gordon Stacey}
\affiliation{Department of Astronomy, Cornell University, Ithaca, NY 14853, USA}

\author[0009-0000-6722-7216]{Thomas Nikola}
\affiliation{Cornell Center for Astrophysics and Planetary Science, Cornell University, Ithaca, NY 14853, USA}

\author[0000-0003-1874-7498]{Cody Lamarche}
\affiliation{Department of Physics, Winona State University, Winona, MN 55987, USA}

\author[0000-0002-1895-0528]{Catie Ball}
\affiliation{Department of Astronomy, Cornell University, Ithaca, NY 14853, USA}

\author[0000-0001-6266-0213]{Carl Ferkinhoff}
\affiliation{Department of Physics, Winona State University, Winona, MN 55987, USA}

\author[0000-0002-4795-419X]{Drew Brisbin}
\affiliation{Joint ALMA Observatory, Alonso de Cordova 3107, Vitacura, Santiago, Chile}

\author[0000-0002-8504-7988]{Steven Hailey-Dunsheath}
\affiliation{California Institute of Technology, Pasadena,
CA 91125, USA}

\newcommand{\tx}{\textrm}

\newcommand{\inv}{$^{-1}$\xspace}
\renewcommand{\epsilon}{\varepsilon}
\renewcommand{\micron}{\textmu m\xspace}
\newcommand{\jykms}{Jy km s\inv}
\newcommand{\msun}{M$_\odot$\xspace}
\newcommand{\lsun}{L$_\odot$\xspace}
\newcommand{\gnot}{G$_\tx0$\xspace}
\newcommand{\dashbetween}{\tx{ -- }}

\newcommand{\aco}{\left(\alpha_{CO}/0.8\right)}
\newcommand{\cii}{[C{\sc ii}]\xspace}
\newcommand{\hii}{H{\sc ii}\xspace}
\newcommand{\nii}{[N{\sc ii}]\xspace}
\newcommand{\niii}{[N{\sc iii}]\xspace}
\newcommand{\oi}{[O{\sc i}]\xspace}
\newcommand{\oiii}{[O{\sc iii}]\xspace}

\ifdefined\garamond
\newcommand{\acro}[1]{\textsc{\lowercase{#1}}}
\newcommand{\acold}[1]{\textsc{\oldstylenums{\lowercase{#1}}}}
\newcommand{\ALMA}{\textsc{Alma}\xspace}
\newcommand{\APEX}{\textsc{Apex}\xspace}
\newcommand{\APECS}{\textsc{apecs}\xspace}
\newcommand{\sdssj}{\oldstylenums{\textsc{sdss j1000}}\xspace}
\newcommand{\jiooo}{\oldstylenums{\textsc{j1000}}\xspace}
\newcommand{\zeus}{\textsc{Zeus-}\oldstylenums{2}\xspace}
\newcommand{\zeusi}{\textsc{Zeus-}\oldstylenums{1}\xspace}

\newcommand{\FUV}{\acro{FUV}\xspace}
\newcommand{\FIR}{\acro{FIR}\xspace}
\newcommand{\ISM}{\acro{ISM}\xspace}

\newcommand{\MCE}{\acro{mce}\xspace}

\newcommand{\PDR}{\acro{PDR}\xspace}
\newcommand{\PDRs}{\acro{PDR}s\xspace}
\newcommand{\PDRT}{\acro{PDRT}\xspace}

\newcommand{\SFR}{\acro{SFR}\xspace}

\newcommand{\TESs}{\acro{TES}s\xspace}
\newcommand{\TES}{\acro{TES}\xspace}

\shortauthors{Rooney et al.}
\shorttitle{Resolved ALMA Study of SDSS J1000}

\begin{abstract}
\shorthandon
We present spatially resolved \ALMA Band-9
observations of the \cii~158~\micron fine structure line from an optically
selected quasar, S*DSS J100038.01+020822.4 (\jiooo), at $z=1.8275$.
By utilizing \oi~63~\micron line observations from Herschel/*PACS and constructing a detailed dust *SED using Herschel and Spitzer archival imaging data,
we show that the \cii line emission is well explained by a photodissociation region (\PDR) model, in which the emission arises from the surfaces of
molecular clouds exposed to far-*UV radiation fields $\sim 5\cdot10^3$ times
the local interstellar radiation field (\gnot). We find a factor of 30 variation in spatially resolved [CII]/Far-IR continuum
across the source which is explained by the reduced fraction of cooling via \cii line emission at such high far-*UV field strengths.
By matching derived \PDR parameters to the observed far-IR line and continuum intensities we derive cloud size-scales and find that typical cloud radii in J1000 are $\sim$ 3.5 pc, perhaps indicating an \ISM that is highly fractured due to intense star formation activity. We model the galaxy dynamically and find that the \cii emission is contained within a compact,
dynamically cold disk with $v/\sigma=6.2$, consistent with the IllustrisTNG50 cosmological simulation. We also report the discovery of a companion galaxy to \jiooo confirmed by the detection of \cii and use recently obtained J\acro{WST}/\acro{NIRC}am
imaging of the system to argue for \jiooo being an interacting system. With total stellar mass $\sim 1.5 \times 10^{10}$ \msun and main-component dynamical mass $\gtrsim 10^{11}$ \msun, the \jiooo system is a progenitor to the most massive galaxies seen in the local Universe.
\shorthandoff
\end{abstract}

\shorthandon
\section{Introduction}
\label{sec:intro}
During the epoch of Cosmic Noon ($z\sim1-3$), the star formation rate density (\acro{SFRD}) peaked, so much so that a majority of stars in the present-day Universe were formed during this time period \citep{Madau2014}. While recent studies have pushed the frontiers of our understanding of galaxy evolution to higher and higher redshifts \citep[e.g.][]{Rybak2020a,Rizzo2022,Atek2023,CurtisLake2023}, constructing a complete picture of galaxy evolution during Cosmic Noon is challenging due to difficulties arising from the fact that important tracers of dusty star formation, such as the far-infrared (\FIR) fine-structure (\acro{FS}) lines \cii 158 \micron and \oiii 88 \micron, are redshifted to narrow, low-transmission telluric windows for ground-based observations.

F*IR *FS lines are an essential part of a complete understanding of the conditions of star formation. Firstly, they are relatively unaffected by dust extinction. This is important since
the majority of the most extreme star-formation episodes and on average about half of the star formation in the history of the Universe occurred in dusty star-forming galaxies (\acro{DSFG}s, see e.g.: \citealp{Houck1985}, \citealp{Whitaker2017}, \citealp{Zavala2021}). Optical wavelength lines will not escape from heavily extincted star formation regions, and even moderate extinction requires correction in line ratios.
The \FIR lines avoid this complexity by simply having longer wavelengths. Second, the excitation potential of \FIR *fs lines, of order 100~K, is very low compared to that of optical lines, and comparable with physical gas temperatures, so \FIR line emissivities are only weakly dependent on gas temperature. In contrast, optical lines, which have emitting levels that are $\sim 30,000$ K above the ground state, have emissivities that are very strongly temperature dependent.

The \cii 158\micron line in particular is among the brightest emission lines from star forming galaxies, often amounting to a few percent of the total luminosity of the system \citep{Carilli2013}, so it is a natural target for studies of galaxy evolution. Since carbon's ionization potential, 11.3 eV, is lower than that of hydrogen, \cii emission can come from the neutral gas phase of the \ISM in addition to the ionized phase. This is useful for dynamical studies, providing velocity information for both of those parts of the galaxy, but complicates the use of \cii as a star formation tracer as \cii comes from these diverse types of regions and its luminosity is sensitive to the physical conditions of the emitting gas. However, in combination with other spectral lines and the far-*IR continuum emission, one can derive the physical conditions of the emitting gas and its cooling rate, and therefore the properties of the sources of gas heating. The preponderance of \cii line emission from most star forming galaxies arises from the warm, dense photodissociated surfaces of molecular clouds exposed to far-*UV (\FUV, $6 \tx{ eV} < h\nu < 13.6 \tx{ eV}$) radiation from nearby OB stars. The \cii line to far-*IR continuum luminosity ratio is a sensitive indicator for the \FUV field strength within these photo-dissociation regions (\PDRs), and by combining it  with other spectral lines and bolometric luminosity measurements, appropriate corrections can be made to measure the physical properties of the gas associated with \PDRs.

{\cii} line emission has been reported from hundreds of extragalactic sources from the initial surveys in the local Universe with the Kuiper Airborne Observatory (e.g. \citealp{Stacey1991}, \citealp{Crawford1985}), *ISO (e.g.  \citealp{Luhman1998}, \citealp{Malhotra2001}) and Herschel (\citealp{DiazSantos2013}), to high redshift surveys with, for example, the Caltech Submillimeter Observatory (*CSO; \citealp{HaileyDunsheath2010}, \citealp{Stacey2010b}, \citealp{Brisbin2015}), \APEX \citep{Gullberg2015,Ferkinhoff2014a}, and the Atacama Large Millimeter/submillimeter Array \citep[\ALMA; e.g.][]{Schaerer2015,Umehata2017,Zanella2018,Schaerer2020,Bouwens2022,Decarli2018,Rybak2021}.

However, the numbers of sources reported from galaxies at Cosmic Noon in the [CII] line emission remains modest due to the challenging nature of observations with the high frequency Bands 9 and 10 of \ALMA \citep{Lamarche2018,McKinney2020,Rybak2021}.
Here we present \ALMA Band 9 observations of the \cii 158~\micron line and the underlying dust continuum from an optically selected quasar, S*DSS J100038.01+020822.4 at redshift $z=1.8275$, hereafter referred to as \sdssj or simply \jiooo. S*DSS \jiooo was selected based on its high 1.2 mm brightness ($4.8 \pm 1.0$ mJy) in the M*AMBO-2/IRAM survey of the C*OSMOS field \citep{Bertoldi2007}. In addition, \cite{Aravena2008} observed four CO lines, (2--1), (4--3), (5--4) and (6--5), with the I*RAM 30 m telescope, and their Large Velocity Gradient analysis indicated that the active galactic nucleus (\acro{AGN}) is likely not the dominant heat source for the molecular gas, instead preferring star formation as the main heat source. S*DSS \jiooo was also included in the first \cii survey at the Cosmic Noon performed using the z (Redshift) and Early Universe Spectrometer (\zeusi) on the Caltech Submillimeter Observatory (*CSO, \citealp{Stacey2010b}), which supported a scenario in which star formation is extended on kiloparsec scales in Cosmic Noon star-forming galaxies and some quasars.

Due to its classification as an optically-selected quasar and its high star-formation rate, \jiooo is referred to as a ``mixed'' system, a type of galaxy only seen in the Cosmic Noon era of the Universe.
In the \cite{Sanders1988} paradigm, this class of systems is transitioning from \acro{ULIRG}-like, with high star formation, to \acro{AGN}-dominated, quiescent systems. This is a reasonable interpretation for \jiooo as shown by the short gas-depletion time observed by \cite{Aravena2008} \citep[for other studies on the interactions between *AGN and star formation see][]{Perna2018,Stacey2018,Valentino2021,FriasCastillo2024}.

The paper is organized as follows.  Section 2 presents our \ALMA observations and ancillary data sets.  Section 3 presents our main results derived from these data, and Section 4 discusses the implications of these results including the confirmation of an associated satellite galaxy and photodissociation region modeling.  Section 5 is our summary and outlook for future investigations.

In this work we assume a flat $\Lambda$ cold dark matter cosmology with $\Omega_m = 0.3$, $\Omega_\Lambda = 0.7$ and $H_0=70 \tx{ km s}^{-1} \tx{ Mpc}^{-1}$. At $z=1.8275$, this corresponds to an angular scale of $8.44\tx{ kpc arcsec}^{-1}$, luminosity distance $D_L = 13.91$ Gpc.

\section{Observations}
\label{sec:obs}
\subsection{\ALMA Observations}
\label{sec:almaobs}
We carried out observations targeting the \cii 158~\micron spectral line and the underlying dust continuum in \sdssj using \ALMA. At $z=1.8275$, the \cii 158~\micron line is redshifted to 446~\micron, placing it in \ALMA Band 9. The observations were carried out in Cycle 3 in two execution blocks on November 16, 2016 and April 22, 2017 under some of the best observing conditions possible at the Chajnantor plateau with precipitable water vapor (\acro{PWV}) 0.26 mm and 0.39 mm respectively.
A total of 42 and 40 antennas were utilized in the 12-m array with source-projected baselines ranging from 15 m to 919 m and 15 m to 430 m in the first and second execution blocks, resulting in maximum recoverable scales of 1\farcs3 and 2\farcs0 respectively.
The
integration time was 22 minutes in each execution block, for a total on-source time of 44 minutes.

We carried out standard data reduction using pipeline version \texttt{42030} within the Common Astronomy Software Application (\acro{CASA}) version \texttt{5.4.0-68} \citep{McMullin2007}. As an additional check, we manually inspected and flagged any outliers and baselines at the raw visibilities level. The data from the two execution blocks was then combined for imaging.

We identified the spectral channels containing the line emission and flagged them before using the *CASA routine \verb|UVContSub| to fit and remove a linear continuum model from the data. We then imaged the continuum-free raw visibilities to create a line cube with a channel width of 50 km/s and pixel size $0\farcs015\times0\farcs015$\
using Briggs weighting with a robustness parameter of 0.5. This results in a synthesized beam size of $0\farcs20\times0\farcs17$ with position angle $55\degree$, which corresponds to a spatial resolution $1.7\times1.4$ kpc at the source redshift. After flagging, the 5th percentile baseline length was 53 m, giving our image a maximum recoverable scale of 1\farcs7. We achieved a 1-$\sigma$ sensitivity of 2 mJy/beam/ch for the spectral line cube.

We also created a dust-continuum image of \sdssj by combining all the line-free spectral channels. The data were imaged with the same pixel size as above using Briggs weighting with a robustness parameter 0.5, resulting in a synthesized beam size of $0\farcs19\times0\farcs16$ with position angle $55\degree$. This corresponds to a spatial resolution of $1.6\times1.4$ kpc. The continuum image has a 1-$\sigma$ sensitivity of 0.3 mJy/beam and is
centered at 669.5 GHz, or 158.3\,\micron rest-frame.

\subsection{Herschel \oi Line Observation}
\label{sec:herschelobs}
To enable modeling of the \cii line emission in a \PDR paradigm, we also utilized \oi 63.184~\micron observations that were conducted with the Herschel/*PACS spectrometer \citep{Poglitsch2010} targeting \sdssj (Program \verb\OT1_gstacey_3\, ObsID \texttt{1342233711}). The redshifted wavelength of $178.56$ \micron was observed in 3 nod cycles with 10 repetitions per nod cycle. The data were reduced with the Herschel Interactive Processing Environment \citep[*HIPE, v12.1.0;][]{Ott2010} with \texttt{oversample=2} for a Nyquist sampled spectral resolution of 103 km s\inv. S*DSS \jiooo is small compared to the beam size of Herschel, so only the spectrum of the central spaxel was considered. After the point-source correction was applied, the spectrum was continuum-subtracted and a Gaussian line profile was estimated from the continuum-subtracted spectrum (\autoref{fig:herschelspec}). The line flux was estimated by integrating the continuum-subtracted spectrum over a wavelength range $\Delta\lambda=2\times$*FWHM of the Gaussian profile. This yields a line flux of $29 \pm 9$ Jy km s\inv including only statistical uncertainty, with an additional 12\% calibration uncertainty\footnote{Herschel Explanatory Supplement Volume III (HERSCHEL-HSC-DOC-2101, version 4.0.1)}.

\begin{figure}
\centering
\includegraphics[width=\columnwidth]{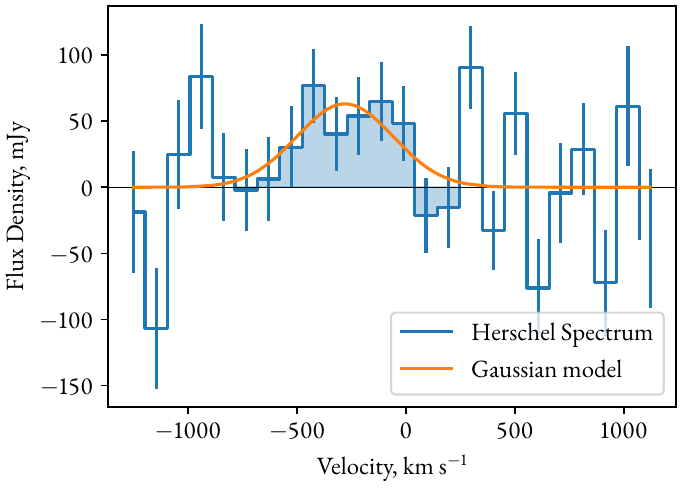}
\caption{Continuum-subtracted Herschel/*PACS spectrum of the \oi 63 \micron line from \sdssj in blue. The shaded region indicates the portion of the spectrum summed to estimate the total line flux, and the orange line shows the best-fit Gaussian line profile. The velocity axis is shown relative to $z=1.8275$ to match the \ALMA spectra presented below.}
\label{fig:herschelspec}
\end{figure}

\subsection{Archival Continuum Observations and Data}
\textsc{Sdss} \jiooo is in the well-studied *COSMOS field~\citep{Scoville2007} and therefore has a rich, multi-wavelength set of archival data.

The newest observations in the archive are from J*WST, and were observed and released on April 15, 2023 as part of the C\acro{OSMOS}-Web Cosmic Origins Survey \citep{Casey2023}, J*WST program number 1727 (PI: Kartaltepe). During this program, J*WST \acro{NIRC}am observed a field containing \jiooo in 4 bands: *F115W, *F150W, *F227W, and *F444W. By mapping the three shorter wavelength bands to red, green, and blue, we created a color photometric image of \jiooo (\autoref{fig:jwst}) and overlaid our \ALMA \cii contours. This extremely deep J*WST imaging reveals possible tidal interaction features between the two sources, and beautiful spiral arms in the main component!

\begin{figure}
\centering
\includegraphics[width=\columnwidth]{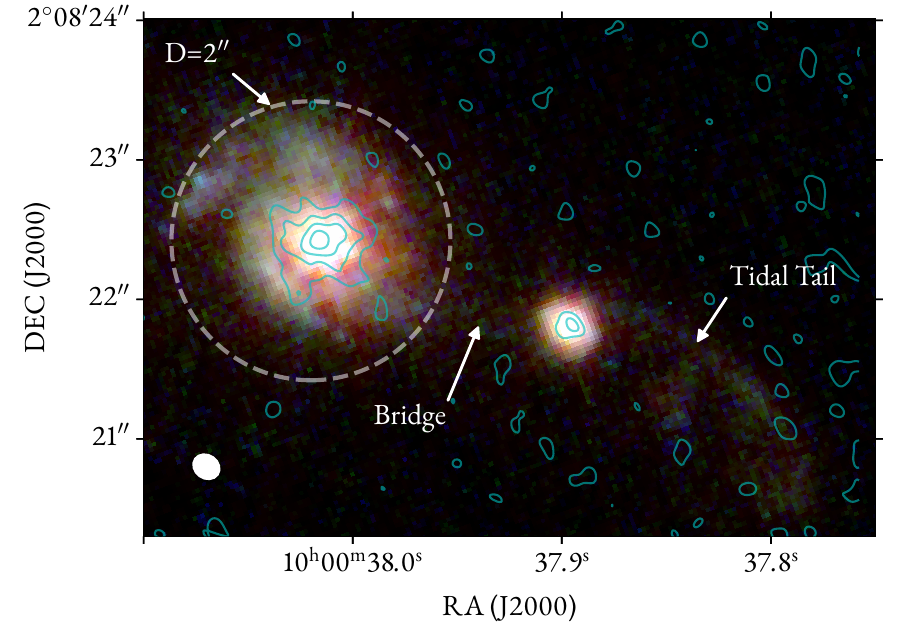}
\caption{A cutout from \acro{JWST} \acro{NIRC}am imaging of \sdssj is displayed with overlayed \cii 158 \micron moment-1 contours (see \autoref{sec:results}) in cyan. The contour levels are multiples of $2\times RMS$, where $RMS= 0.7$ Jy km s\inv beam\inv. N\acro{IRC}am images from three bandpass filters, \acold{F115W}, \acold{F150W}, and \acold{F277W} were mapped to the blue, green, and red channels (respectively) of the displayed cutout. The astrometric calibration of the \acro{JWST} data is good to within $\sim 0\farcs1$ . The possible tidal tail and bridge between the two components are called out. The 2\arcsec\ diameter circle is provided to give an idea of the scale of the emission relative to the \ALMA maximum recoverable scale, which is 2\arcsec\ in the most optimistic scenaro.
}
\label{fig:jwst}

\end{figure}

We extracted far-*IR to sub-mm photometry for \sdssj from the Herschel Point Source Archive \citep{Marton2017,Schulz2017} for both *SPIRE and *PACS, yielding five photometric data points between 100-500 \micron (see \autoref{tab:sed}) sampling the peak of the dust emission. The Herschel beam size ranged from 7\arcsec\ for the 100 \micron observation to 30\arcsec\ for the 500 \micron observation, so no morphological information could be recovered. Uncertainties due to confusion are included in the errors reported in \autoref{tab:sed}.

Finally, we used (observed frame) 24 and 70 \micron fluxes from Spitzer/*MIPS for *SED modeling, from \cite{Sanders2007}, 872.6 \micron flux from \ALMA program \texttt{2016.1.00463.S}, 2068 \micron flux from \ALMA program \texttt{2021.1.00705.S}, and 3038 \micron flux from \ALMA program \texttt{2021.1.00246.S} (see \autoref{tab:sed}).

\section{Results}
\label{sec:results}
Our \ALMA observations detect the \cii 158 \micron line and continuum from \sdssj at high significance, as seen in the spectral line map with continuum contours in \autoref{fig:images} (top). The moment-0 map in \autoref{fig:images} (top) was constructed by stacking the channels containing the line, resulting in an image with an $RMS = 0.7$ Jy km s\inv beam\inv. The moment-1 map also shows \cii 158 \micron flux co-spatial to a source previously reported in *HST imaging by \cite{Aravena2008} $1\farcs9$ west-south-west (\acro{WSW}) from the main source.

To estimate line and continuum fluxes, we numerically integrate the flux within a circular aperture centered on the source, and to estimate the error in these flux measurements we measure the *RMS in an equal-sized aperture centered away from the source (in units of Jy/beam) and multiply by $\sqrt{n_{beams}}$. For the main source's \cii flux, we use a diameter $=1\farcs45$ and calculate a total line flux of $37.7 \pm 3.3$ \jykms.
This flux is about half the flux detected by \acro{CSO}/\zeusi observations conducted by~\cite{Stacey2010b}.
Much of this could be attributed to flux calibration and statistical errors within the \zeusi observations, but in light of the large scale stellar emission detected with J*WST \acro{NIRC}am (see \autoref{fig:jwst}) it is also likely that some \cii line flux observed with the $11\farcs5$ \zeusi beam is ``resolved out'' by \ALMA (see \autoref{sec:resolvedout}).

A bright source to the west of \sdssj is detected in H*ST/ACS imaging \citep{Aravena2008}. We report the detection of \cii 158 \micron line emission co-spatial with this source, at the same redshift, thereby spectroscopically confirming that it is physically associated with \jiooo. We will refer to \jiooo as the ``main'' source and the western companion source as the ``satellite'' in this paper. We extract the \cii flux for the satellite using a circular aperture of $0\farcs35$ and detect it at $4\sigma$ significance with a flux of $2.3 \pm 0.6$ \jykms.

The dust continuum emission from the main source is spatially resolved but is found to be relatively compact as compared to the extent of the \cii line emission: The 2-D Gaussians of best fit to the \acro{CLEAN}ed images have \acro{FWHM}s $0\farcs216 \pm 0\farcs006$ for the continuum and $0\farcs43 \pm 0\farcs04$ for the line emission. There is extended emission beyond the Gaussian shape, however, for both the continuum and the line, so we use larger diameter apertures for computing the total flux of both. Using a circular aperture of $1\farcs05$, we measure the main source's continuum flux density $=55.6 \pm 1.8$ mJy.
We also measure the integrated flux density for the satellite in a $0\farcs4$ diameter aperture to be $3.5 \pm 0.7$ mJy.

This \ALMA data has considerably higher spectral resolution than previous observations of this source. Using the same apertures as we used for the \cii line integrated flux measurements, we also create the spectral profiles seen in \autoref{fig:spectra}. We clearly recover a double-horned spectral profile for \jiooo offset by $+80$ km s\inv relative to the previously reported redshift of $z=1.8275$ and report a \cii redshift of $z=1.82825\pm0.00002$.

The double-horned spectral profile and the symmetry seen in the moment 1 map (\autoref{fig:images}, bottom) suggest that a rotationally supported disk geometry is an appropriate model for the dynamical structure of \jiooo. We explore this idea further in \autoref{sec:disk}.

Leveraging the sensitivity of our continuum and spectral line maps, we identify a spatial offset of $0\farcs04 \pm 0\farcs02$ between the dust continuum peak and the peak of \cii line emission. This modestly significant offset can be seen by eye by comparing the contours to the image in \autoref{fig:images} (top).

\begin{figure}
\centering
\includegraphics[width=\columnwidth]{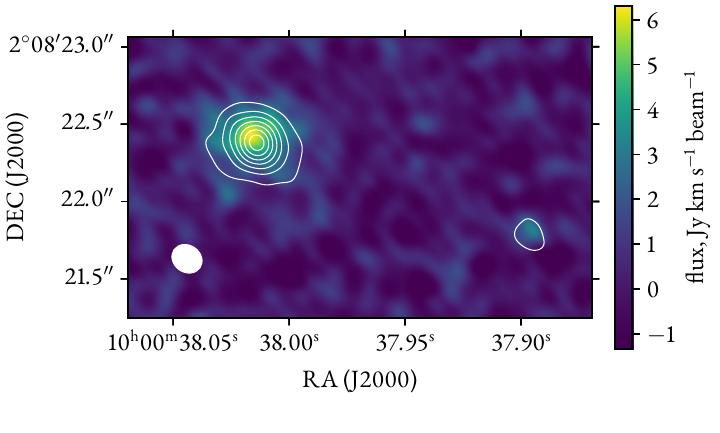}
\includegraphics[width=\columnwidth]{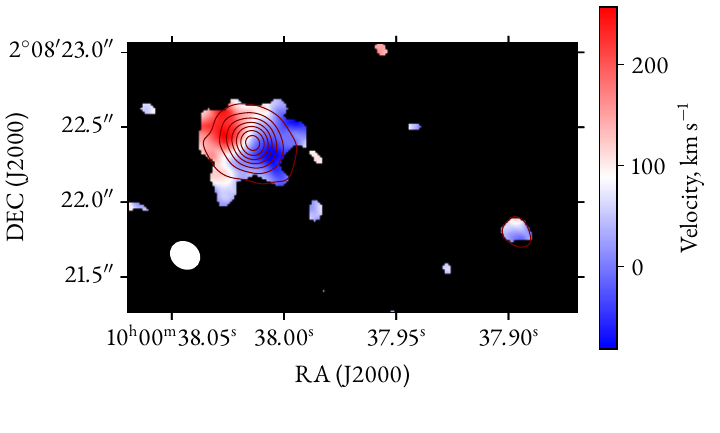}
\caption{Top: The moment-0 map of the \cii line in \sdssj, with rest-frame 158\micron continuum contours superimposed. Bottom: The moment-1 velocity map of the \cii line emission, also with rest-frame 158\micron continuum contours. In both cases the continuum contour levels are 1.5--19.5 mJy in steps of 3 mJy.}
\label{fig:images}
\end{figure}

\begin{figure}
\centering
\includegraphics[width=0.8\columnwidth]{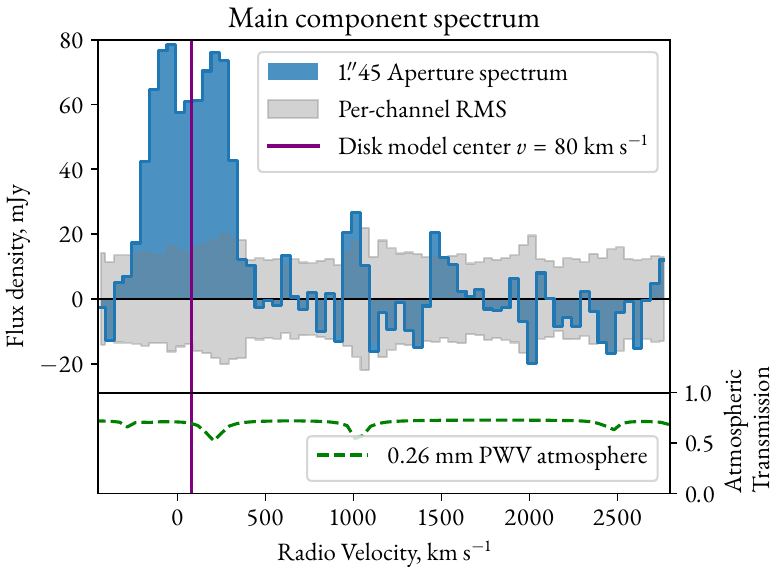}
\includegraphics[width=0.8\columnwidth]{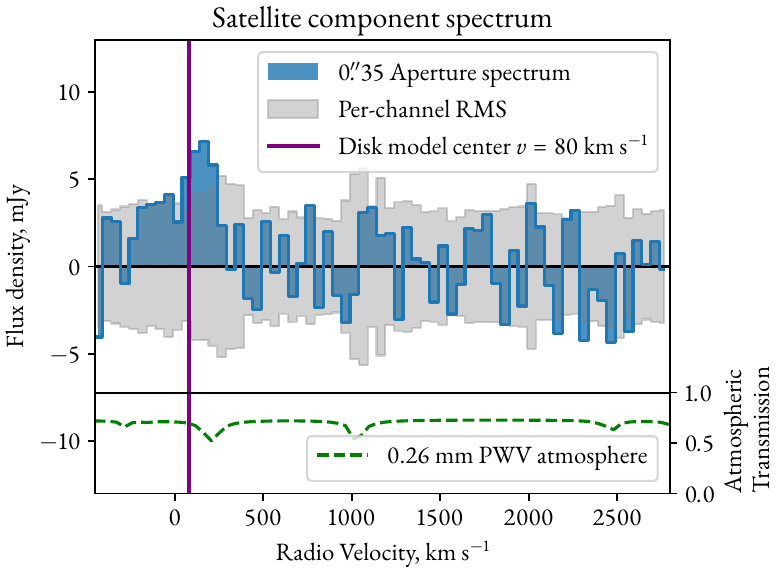}
\caption{The continuum-subtracted \cii 158 \micron spectrum of \sdssj's components: extracted from an aperture of diameter $1\farcs45$ centered on the main source (top) and, extracted using an $0\farcs35$ aperture centered on the satellite source (bottom). Velocity is plotted with respect to $z=1.8275$, with the vertical line indicating the central velocity from the disk modeling routine. The atmospheric transmission is plotted below the spectrum.}
\label{fig:spectra}
\end{figure}

\subsection{SED analysis}
\label{sec:sed}
Using publicly-available Herschel data and previously-published continuum measurements (listed in \autoref{tab:sed}), we create a model for the dust *SED of \sdssj (\autoref{fig:sed}). We use the modified blackbody routine {\sc mercurius}~\citep{Witstok2022}, which parameterizes the dust emission in the form
\begin{equation}
S_\nu = M_{\tx{dust}}\;\kappa_\nu \;\frac{1+z}{D_L^2(z)} \;\left[B_\nu\left(T_{\tx{dust}}\right) - B_\nu\left(T_{\tx{CMB}}\left(z\right)\right) \right]
\end{equation}
where $\kappa_\nu\propto\nu\,^\beta$, to model the continuum emission from \sdssj. With this model, we find \FIR\footnote{Following \cite{Brisbin2015}, we define \FIR luminosity as the integrated luminosity between 42.5 and 122.5 \micron, and *TIR (total infrared) luminosity as the integrated luminosity from 8-1000 \micron} luminosity $=(7.2 \pm 0.3)\cdot10^{12}$ \lsun , star-formation rate $=1640\pm80$ M$_\odot$ yr\inv, and dust mass $=(1.8\pm0.1)\cdot10^9$ M$_\odot$. These results match those obtained through
the methods detailed in \cite{Brisbin2015}, using template fitting of \cite{Dale2002} models.

\begin{table}
\centering
\caption{S*DSS *J1000 continuum fluxes
used for *SED modeling. Spitzer *MIPS
data are provided by the *S-COSMOS project \cite{Sanders2007}; *BOLOCAM data is from \cite{Aravena2008}. Herschel/*PACS and *SPIRE data from their respective Herschel Point Source Catalogs: \cite{Marton2017} and \cite{Schulz2017}. All of these observations treat \jiooo as a point source. }
\label{tab:sed}
\begin{tabular}{|r|cc|}
\hline
\parbox[t]{1.5cm}{\centering $\lambda_{obs}$ [\textmu m]}& Flux Density [mJy]             & Source \\
\hline
70    &$   7      \pm  2      $& *mips \\
100    &$  24      \pm  4      $& Herschel/*PACS \\
160    &$  73      \pm  9      $& Herschel/*PACS \\
250    &$  85      \pm  5      $& Herschel/*SPIRE \\
350    &$  69      \pm  8      $& Herschel/*SPIRE \\
500    &$  35      \pm  7      $& Herschel/*SPIRE \\
446    &$  55.6    \pm  1.8    $& \ALMA/this work \\
873    &$  9.8     \pm  0.4    $& \ALMA \texttt{2016.1.00463.S} \\
1100    &$   5.6    \pm  1.9    $& Bolocam \\
1200    &$   4.8    \pm  1.0    $& M*AMBO \\
2068    &$  0.3     \pm   0.1   $& \ALMA \texttt{2021.1.00705.S} \\
3038    &$  0.11    \pm   0.02  $& \ALMA \texttt{2021.1.00246.S} \\
\hline
\end{tabular}
\end{table}

\begin{figure}
\centering
\includegraphics[width=\columnwidth]{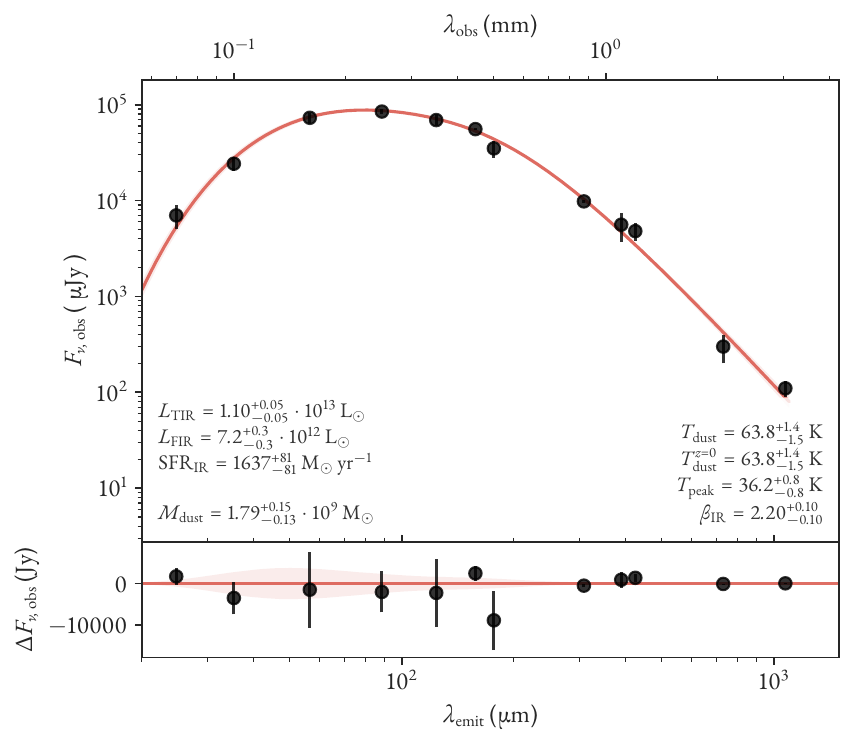}
\caption{Our compiled *SED of \jiooo spans the whole dust continuum, allowing tight constraints on the temperature of the dust and the derivation of a very accurate $L_{FIR}$. Raw data for this plot is given in \autoref{tab:sed}. The model here assumes $\lambda_0=250$ \micron. The 446~\micron point is from this work.
}
\label{fig:sed}
\end{figure}

\subsection{Disk fitting}
\label{sec:disk}

As mentioned above, the symmetry in the moment-1 map of \sdssj suggests the presence of ordered rotation. To investigate this further, we use the UVModelDisk routine from \cite{Pavesi2018} to model \jiooo as a rotating disk. Within UVModelDisk, the goodness-of-fit parameter is computed by transforming the disk model to the *UV plane and comparing the resulting visibilities to the calibrated \ALMA visibilities. This is beneficial due to the fact that visibilities are statistically independent, while pixels in \ALMA images are not, and thus no covariance matrices need to be computed when running UVModelDisk. More technical details of the process are described in \autoref{sec:disktech}.

\subsubsection{Model parameters}

Ten physical parameters of a rotating disk, listed in \autoref{tab:modelparams}, were varied during model construction, and an additional four parameters corresponding to the continuum emission were fixed to the values output by the *CASA routine \verb\UVModelFit\ in ``Gaussian Source'' mode on channels which do not contain the line. This produces a 2-dimensional Gaussian model of the continuum in the image plane, and assumes that the continuum is constant in frequency. For each parameter that is allowed to vary, UVModelDisk returns its posterior distribution and its covariance with each other parameter.
\shorthandoff

\begin{table*}

\caption{[Top] The parameters used in the UVModelDisk routine. Parameter names are listed in the first column, followed by the prior ranges used in the nested sampling routine, the prior types used, and the median and $1\sigma$ width of the posterior distribution. [Bottom] Parameters for the continuum model were fixed beforehand and not allowed to vary during the sampling.}
\label{tab:modelparams}
\shorthandon

\begin{tabular}{|c|ccc|}
\hline
Parameter name & Prior & Prior type & Posterior \\
\hline

Gas velocity dispersion ($\rho_{\sigma}$) [km/s] & 10-700 & Log & $56.8^{+2.6}_{-2.2}$\\
Brightness *FWHM [arcsec] & 0.05-1  & Log & $0.456^{+0.012}_{-0.011}$\\
Maximum Rotational Velocity [km s$^{-1}$] &100-1000&Log &$351^{+25}_{-19}$\\
Velocity Scale [arcsec] & 0.001-1 & Log & $0.001^{+0.002}_{-0.001}$ \\
Velocity center [GHz, observed]& 671.8-672.2& Uniform & $671.9833^{+0.0047}_{-0.0058}$\\
Inclination angle [deg] & 20-70 & Sin & $36.5^{+2.6}_{-2.9}$\\
Position angle [deg] & 10-100 & Uniform &$28.9\pm1.1$\\
R*A center [arcsec]$^\tx{a}$ &-0.1-0.1&Uniform& $0.03\pm0.003$\\
D*EC center [arcsec]$^\tx{a}$ &-0.1-0.1&Uniform& $-0.001\pm0.003$\\
Total Flux [Jy km s$^{-1}$] &25-35&Log&$28.1^{+0.8}_{-0.7}$ \\
\hline
Continuum *RA offset [arcsec] & 0 & [fixed] & \\
Continuum *DEC offset [arcsec] & 0 & [fixed] &\\
Continuum flux *FWHM [arcsec] & 0.22 & [fixed] & \\
Continuum flux density [mJy] & 52.7 & [fixed] & \\
\hline

\end{tabular}

\shorthandoff

\tablenotetext{a}{offset from continuum center}

\end{table*}

\shorthandon

\subsubsection{Model Results}
\label{sec:modelresults}
The posteriors of each parameter are given in \autoref{tab:modelparams}.
The largest source of error in the maximum rotational velocity and inclination angle is their covariance due to geometrical projection, and the velocity scale is not well-constrained because it is much smaller than the resolution of the observations. The other parameters are relatively well-constrained and do not depend strongly on each other.
Images of the best model and its residuals are discussed in \autoref{sec:residuals}.

Of particular note are the values obtained for the velocity dispersion of the gas, $\rho_{\sigma} = 56.8^{+2.6}_{-2.2}$ km s\inv, and the maximum velocity of the rotation curve, $v_{max} = 351^{+25}_{-19}$~km~s\inv. This value is already corrected for inclination $\alpha = 36.5^{+2.6}_{-2.9}$ and assumes a standard arctan velocity curve
\begin{equation}
v(r) = \frac{v_{max}}{\pi^2} \arctan\left(\frac{r}{r_{0,v}}\right)
\end{equation}
Here, $r_{0,v}$ is the characteristic distance scale of the velocity curve, expressed in terms of apparent distance on sky in arcsec. In our case, where $r_{0,v} \ll$ beam size, this reduces to a flat rotation curve where $v=v_{max}$.

Following \cite{Rizzo2020}, we can compute the $v/\sigma$ ratio to determine how ordered or disordered the galaxy's bulk motion is. When $v/\sigma > 2$, ordered motion dominates over random motion. We calculate $v/\sigma = 6.2$ for \jiooo, which indicates that it is a dynamically cold disk. This places it squarely within the range predicted by the Illustris *TNG50 simulation \citep{Pillepich2019} in terms of dynamical evolution as presented in Fig. 3 of \cite{Rizzo2020}. However, other galaxies are presented there too which are cold disks at much earlier times ($z\sim4$).

\subsection{PDR model}
\label{sec:pdr}
Assuming that the \cii and \oi lines and \FIR continuum arise from \PDRs, we use the \textsc{Pdr} Toolbox \citep[\PDRT;][]{Kaufman2006,Pound2008,Pound2011,Pound2023} to estimate the \FIR fine-structure line intensities produced by gas with density $n$ and incident far-*UV flux\footnote{Following \cite{Tielens1985a} we define $G_0$ as the Habing flux ($G_0 = 1.6\times 10^{-6} \textrm{ W m}^{-2}$ or $1.3\times 10^{-7} \textrm{ W m}^{-2}\textrm{ sr}^{-1}$ in intensity units) and express the far-*UV flux $G$ in terms of $G_0$.} $G$.
The \verb\wk2020\ models included in \PDRT make calculations to a depth equal to visual extinction A$_V=7$ and Solar metallicity (Z~$=1$), and the model \FUV field strengths and gas densities are plotted as a grid in \autoref{fig:oiciifir} as functions of two diagnostic line ratios: \cii 158 \micron/\oi 63 \micron and (\oi 63 \micron + \cii 158~\micron)/\FIR.

To compare this model with our observations, we start with our observed \cii luminosity, our derived $L_{FIR}$ from \acro{SED} modeling, and the \oi 63 \micron luminosity~$=(9.8\pm3.7)\cdot10^9$~L$_\odot$, and apply corrections for ionized gas and optical depth. About 25\% of the observed \cii line radiation likely arises from ionized gas \citep{Oberst2006}, so we multiply the \cii line luminosity by a factor of $3/4$. If the emitting clouds are externally heated spherical clouds, the observed optically thin \cii line and \FIR continuum will arrive from the entire clouds' surfaces, with area $4\pi r^{2}$, while the optically thick \oi 63 \micron line will only arrive from a surface of area $\sim\pi r^{2}$. We therefore multiply the observed \oi line by a factor of 2 since something between a factor of 1~and 4~of the \oi line flux will be blocked from our view. After applying these corrections, we plot the diagnostic line ratios for \jiooo and all other galaxies with the same line observations available in the literature in \autoref{fig:oiciifir}.

In \autoref{fig:spaghetti} we show a different view of the \PDRT model for \sdssj, where the parameter space of $G$ and $n$ is shown shaded with the allowed regions for different line ratios. The solution at $n=6\times 10^3\textrm{ cm}^{-3}$, and *UV field $G=5\times 10^3\;G_0$ corresponds with the value shown in \autoref{fig:oiciifir}. Focusing only on \oi, \cii, and \FIR, another solution also exists, with $n=1.5\times 10^5\textrm{ cm}^{-3}$, $G=3\;G_0$, and distinguishing between the two involves thoughts about filling factors. Dust is heated by nearby starlight and re-radiates its energy in the \FIR continuum. Typically half the starlight that heats this dust is in the \FUV, so that a beam filling factor of unity would have \FIR intensity matching half the \FUV intensity:  $I_{FIR} = L_{FIR}/(4\pi d^2 \Omega_{source})=1.9\times 10^{-3} $ W m$^{2}$ sr$^{-1} = 2G \Rightarrow G= 1.5\times 10^{4} G_{0}$. The low $G$ solution requires a filling factor of 5000, which is implausible. The high $G$ solution has a much more plausible beam filling factor of 4.

Looking at the CO lines plotted in \autoref{fig:spaghetti} paints a somewhat different picture. The ratios of different CO transitions agree with the \oi, \cii, and \FIR solution to within a couple $\sigma$, but the \cii/CO ratio intersects with the others at a very different point in parameter space, which indicates an excess of CO line emission. This excess could be explained in a couple of different ways. In the standard \PDR paradigm (which is what we adopt for the remainder of this work), \oi, \cii, and \FIR emission are predominantly from the \PDR itself, while the CO emission could arise from molecular clouds not associated with \PDRs in addition to those which are \citep{Mashian2015} or could additionally be excited by shocks. Alternatively, the \oi or CO emission could arise from \acro{XDR}s in addition to \PDRs \citep{vanderWerf2010}. We adopt the explanation that the CO emission comes from cooler, higher-density parts of the molecular clouds, but distinguishing between these scenarios requires additional observations.

\begin{figure}
\includegraphics[width=\columnwidth]{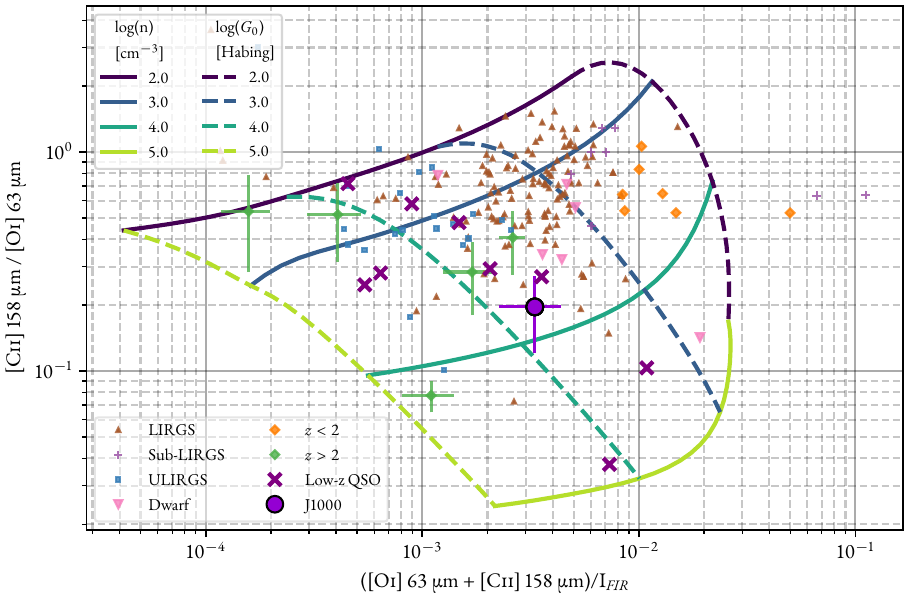}

\caption{Here we plot the \PDR parameter space $(G$, $n)$ from \cite{Pound2023} for two \FIR line diagnostics, the \cii 158 \micron/\oi 63 \micron ratio and \oi 63 \micron + \cii 158 \micron to \FIR ratio. The two spectral lines account for most of the gas cooling. For comparison, we plot observations of \jiooo (purple) and all other galaxies with \cii, \oi, and \FIR observations that we could reasonably account for in the literature: \cite{Zhang2018,FernandezAranda2024,Rigopoulou2018,Brisbin2015,Zhao2016,DiazSantos2017,Brauher2008}; and \cite{Ishii2024}. Line fluxes for the plotted galaxies have been corrected for ionized gas and optical depth by multiplying the \cii flux by 0.75 and the \oi flux by 2 (see text of \autoref{sec:pdr}). Representative uncertainties can be assumed to be 0.3 dex for points not presented with error bars. }
\label{fig:oiciifir}
\end{figure}

\begin{figure}
\centering
\includegraphics[width=0.8\columnwidth]{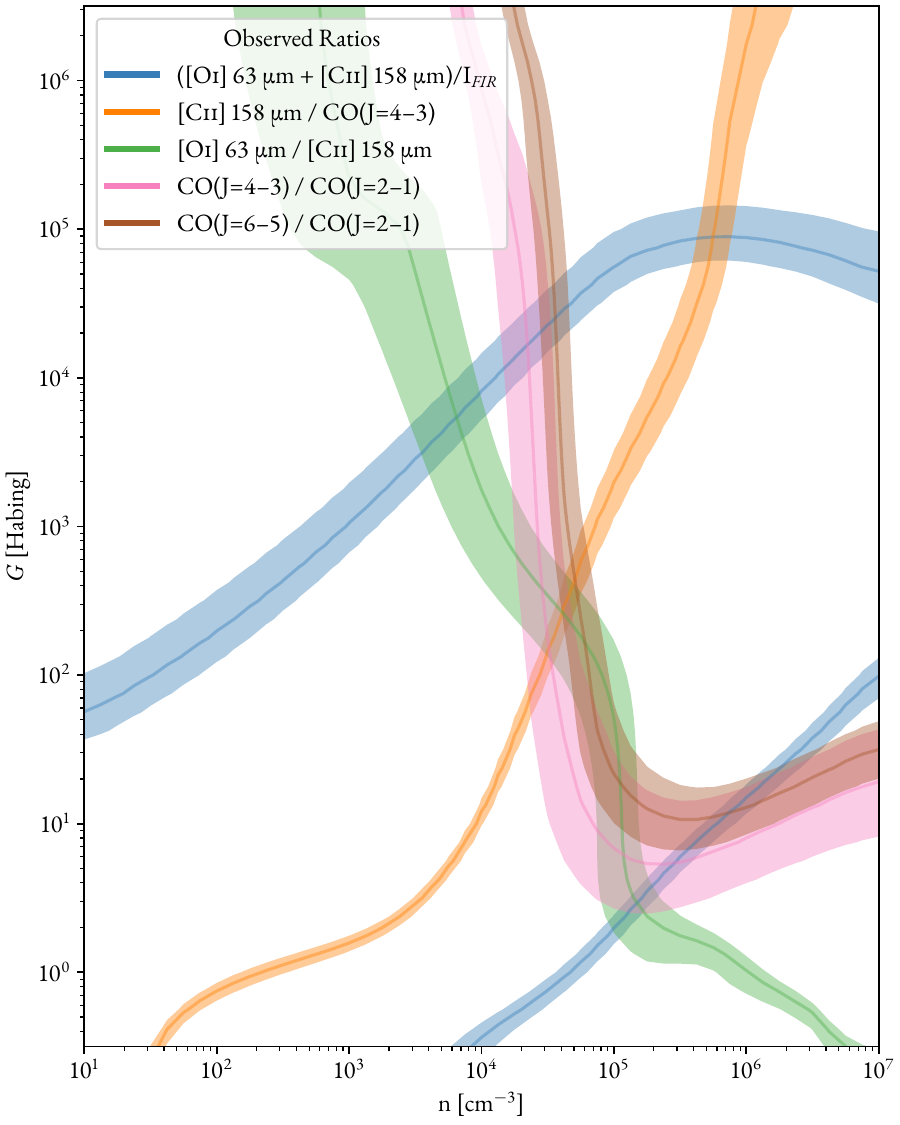}
\caption{P*DRT \citep{Pound2023} model showing regions in $(G$, $n)$ parameter space allowed by ratios of the two fine-structure line luminosities and $L_{FIR}$ we have observed, and CO luminosity from \cite{Aravena2008}. The disagreement between ratios involving CO (pink, brown, orange) and those involving only fine-structure lines and $L_{FIR}$ (blue and green) could have several explanations (see text), but distinguishing between them would require resolved observations of CO in this source. In this paper we adopt the solution at $n=6\times 10^3\textrm{ cm}^{-3}$, and $G=5\times 10^3\;G_0$, with uncertainties $\sim 0.3$ dex.}
\label{fig:spaghetti}
\end{figure}

\subsection{\cii/FIR Map}

Our spatially-resolved observations of \sdssj allow us to produce a map of the \cii/\FIR ratio. By assuming a constant dust temperature throughout the source,
we create a map of the spatial distribution of the total \FIR luminosity of the galaxy. Next, we combine this with our \cii luminosity map to produce the ratio map. We mask the image by including only pixels that lie in the contiguous regions around the two sources where the \cii luminosity and \FIR luminosity remain above $1.3\cdot10^8$ L$_\odot$/px and $5\cdot10^{10}$~L$_\odot$/px respectively, which correspond to 1$\sigma$ thresholds. Using our values from \PDRT and assuming a constant density $n=10^4$~cm$^{-3}$ we produce the map of far-\acro{UV} flux incident on the galaxy's \PDRs shown in \autoref{fig:ratiomap}.

\begin{figure}
\includegraphics[width=\columnwidth]{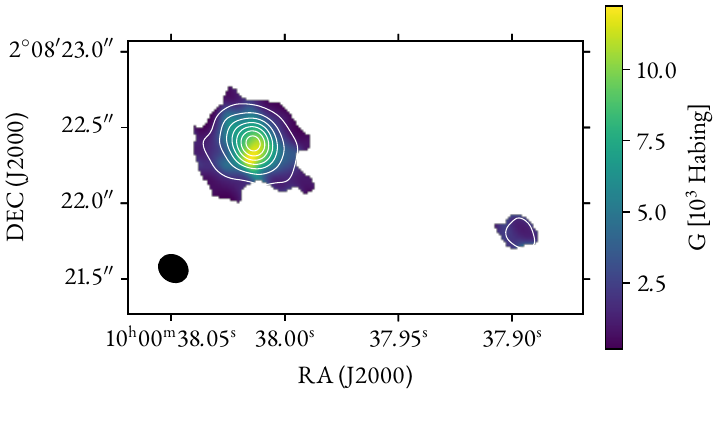}

\caption{Spatially resolved map of the far-\acro{UV} field strength $G$ in \sdssj, derived from the \cii/\FIR ratio using the \cite{Pound2023} models. We assume a constant gas density of $n=10^4$ cm$^{-3}$ throughout the galaxy, but this choice has little effect on the plot given that the \cii/\FIR ratio is insensitive to density \citep{Stacey2010b}. Superimposed are rest-frame 158 \micron continuum contours, the same as \autoref{fig:images}. The \ALMA beam size is shown in the lower-left corner in black. Values of $G$ correspond inversely to the raw value of the \cii/\FIR ratio as shown by the solid blue line in \autoref{fig:sigmasfr}. }
\label{fig:ratiomap}

\end{figure}

\begin{figure}
\includegraphics[width=\columnwidth]{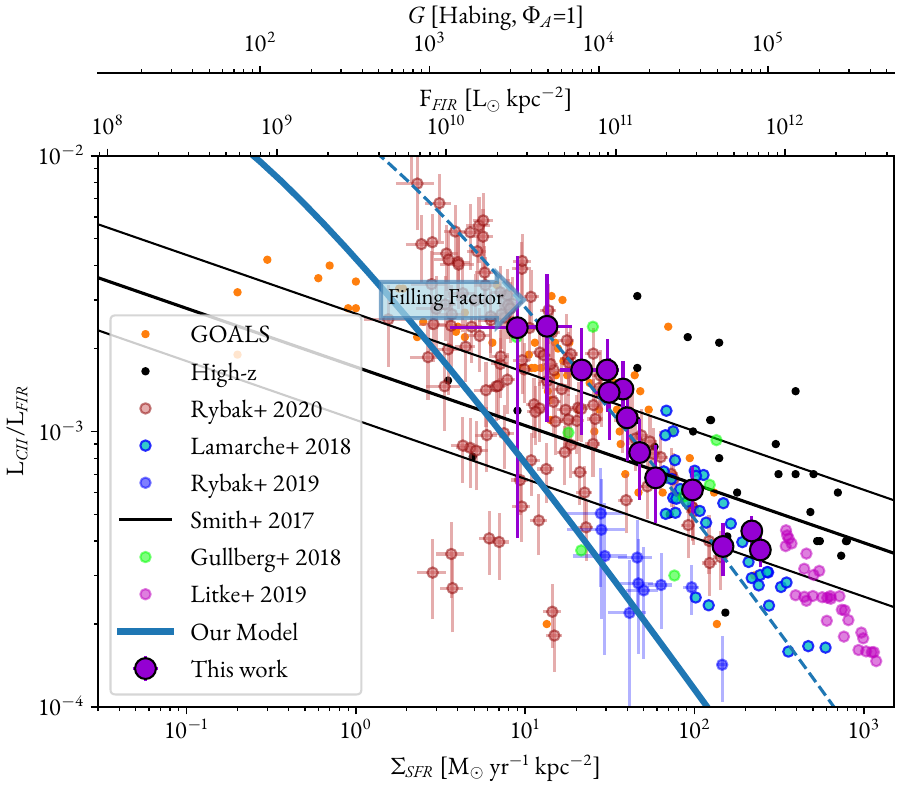}

\caption{Resolved \cii/\FIR measurements in \sdssj (large violet circles) are compared to similar resolved measurements of other high-z targets from \cite{Rybak2020b,Lamarche2018,Rybak2019,Gullberg2018,Litke2019}, and unresolved measurements from \acro{GOALS} \citep{DiazSantos2017} and unresolved high-z observations presented in~\cite{Smith2017}. Each data point presented here for \sdssj represents one independent beam from our \ALMA data. Also shown are the corresponding \acro{FIR} flux values and far-\acro{UV} flux values to the \acro{SFR} surface density on the x-axis. \FIR is scaled to $\Sigma_{SFR}$ using the relationship of \cite{Kennicutt1998} and to $G$ using the relationship of \cite{Kaufman1999}. The model in solid blue is the \PDR toolbox model from \cite{Pound2023}, relating $G$ to \cii/\FIR asuming $n=10^4\tx{ cm}^{-3}$ and the dotted line is the same model shifted by the filling factor $\phi_A=4.2$.
}
\label{fig:sigmasfr}
\end{figure}

Much has been said about the ``\cii deficit'', where the \cii line luminosity decreases as a function of far-\acro{IR} luminosity, and a variety of models, some rather complex, have been proposed to explain the deficit. The simplest way to understand the \cii deficit, however, is by the physics of \PDRs: simply put, the efficiency of line production is reduced as the strength of the \FUV field goes up.  As we will show, our spatially resolved observations of the \cii/\FIR ratio (\autoref{fig:sigmasfr}) are in clear agreement with this idea.

In \autoref{fig:sigmasfr}, we plot the \cii/\FIR ratio in the various beams across \sdssj against the far-IR flux scaled to \SFR surface density $\Sigma_{SFR}$ using the relationship of \cite{Kennicutt1998}:
\[
\tx{SFR [M}_\odot\tx{ yr}^{-1}\tx{]} = 1.7\cdot10^{-10} \tx{ L}_\tx{FIR} \tx{[L}_\odot\tx{]}
\]
Alongside our resolved measurements we plot a blue line showing the predicted \cii/\FIR for a given $G$ from \cite{Pound2023}, scaled to \FIR by the relationship in \cite{Kaufman1999}. It is immediately apparent that the slope of the theoretical line matches the slope of the observed ratios from many resolved galaxies. More quantitatively, \cite{Rybak2020b} and \cite{Lamarche2018} determined power-law slopes of $-0.75$ and $-0.7$ for \acold{SDP.81} and \acold{SDP.11} respectively. We find that \sdssj has a power-law slope of $-0.61 \pm 0.08$, and \cite{Pound2023}'s \PDR model grids (which assume a plane-parallel geometry) have a power-law slope of $-0.80 \pm 0.05$, indicating that \jiooo and the other resolved galaxies have a \cii \emph{excess}, not deficit. The horizontal offset between the model value and the observed relationship is simply the area filling factor of the source, since increasing filling factor increases \FIR luminosity without changing any ratios. An adjustment for the filling factor ($\phi_A=4.2$, dotted blue line) shows excellent agreement with the data, making a compelling argument that the \cii emission is dominated by \PDRs, and the dependence of \cii emission on \FIR flux is explained simply by the physics of the \PDR. There are two effects in the details of \PDR models that predict a reduction in the \cii to \FIR continuum luminosity ratio as the strength of the \FUV field grows. First, in the regime of very intense \FUV fields, in particular when $G/n$ is large, the large charge on a grain will increase so that the energy lost by a photo-electron ejected from such a grain decreases, thereby decreasing the gas heating rate. Second, since the penetration of \FUV photons capable of ionizing carbon to form C+ is limited by dust extinction, the
emitting column of C+ grows logarithmically with \FUV flux. In contrast, the optically thin dust emission grows linearly with \FUV so that the line to continuum ratio decreases with increasing \FUV field strength
\citep{Wolfire1990, Stacey1991, Stacey2010b}.
The excellent fit of the data to \PDR models is very strong evidence that the so-called ``\cii deficit'' is the expected result of standard \PDR physics.

The differences in slope from this work to \cite{Smith2017} and \cite{DiazSantos2013} in resolved star-forming galaxies and local \acro{ULIRG}s respectively can likely be attributed to geometry. \cite{Ferkinhoff2015a} and \cite{HerreraCamus2018b} show that the expected dependence of \cii/\FIR on $\Sigma_{\tx{SFR}}$ changes based on geometry and total $\Sigma_{\tx{SFR}}$. In geometries with \PDRs closely associated to individual ionizing sources, the power-law slope is consistent with that of local \acro{ULIRG}s at $-0.26$, but this relation steepens at $\Sigma_{\tx{FIR}} > 3\cdot10^{10}$ L$_\odot$ kpc$^{-2}$ or when the geometry is that of a uniform interstellar radiation field irradiating molecular clouds with \PDRs at the interface. It is conceivable that galaxies with more extreme $\Sigma_{\tx{FIR}}$ are more likely to take on the second geometry as \acro{UV} fields become stronger throughout the source.

\section{Discussion}
\subsection{Extended flux: Resolved Out Emission or Low Surface Brightness?}
\label{sec:resolvedout}
The \zeusi detection of \cii from \sdssj \citep{Stacey2010b} found significantly higher line flux than detected in our \ALMA imaging ($78\pm18$ Jy km s\inv and $37.7\pm3.3$ Jy km s\inv, respectively). The values are nearly consistent if we take into account calibration uncertainties in the \zeusi data and its modest statistical significance.  If we take the flux discrepancy at face value, it could also be explained by the presence of a diffuse component to the \cii emission that is detected within the $11\farcs5$ \zeusi/*CSO beam but either resolved out by the maximum recoverable scale of our \ALMA observations or at a surface brightness that is simply too low to be recovered from the noise in the \ALMA map.

Using new observational constraints from \acro{NIRC}am imaging, we can estimate the expected surface brightness of the ``missing'' flux. It is reasonable to assume that the gas brightness distribution is traced by rest-frame stellar emission seen in the \acro{NIRC}am imaging, which is clearly distributed on scales of $\sim 2\arcsec$. If we assume the extra flux detected by \zeusi is spread uniformly on a $2\arcsec$ diameter disk (84 beams), the expected surface brightness would be 0.3 Jy km s\inv beam\inv. This is only 40\% of the \acro{RMS} value of the line image: 0.7 Jy km s\inv beam\inv.

This already low detectability is further compounded by the maximum recoverable scale (\acro{MRS}) of our \ALMA observations. The \acro{MRS} is the size of a uniform disk for which observations would detect 10\% of the flux (\ALMA technical handbook cycle 4), and response scales as $\propto\exp(\theta^2 L_\tx{min}^2)$ \citep[Where $L_\tx{min}$ is the minimum baseline length;][]{Wilner1994}. As mentioned in \autoref{sec:almaobs} the \acro{MRS}s for our observations were 1\farcs3 and 2\farcs0 in the two execution blocks, and in the final combined image it was 1\farcs7. So in the most optimistic scenario with \acro{MRS}=2\farcs0, much of any putative emission at $2\arcsec$ scales would be resolved out. For comparison, a $2\arcsec$ circle is drawn in \autoref{fig:jwst}.

\subsection{Confirmation of Physically Associated Satellite}

We confirmed the redshift of the satellite galaxy, detecting its \cii line at redshift $z = 1.82821 \pm 0.00040$, consistent with the center velocity of the main component
($1.82825 \pm 0.00002$). This indicates the two components are physically associated, and allows us to estimate a merger timescale. If we assume the line of sight difference between the two sources is small and calculate $t_{dyn}$ using only the transverse distance, $15.7$ kpc, and the mass of the system derived by \cite{Liu2019}, $1.5\cdot10^{11}$~M$_\odot$, we calculate $t_{dyn}\sim400$ Myr. Merger timescales are estimated to be similar to the dynamical timescales within a factor of a few \citep{Solanes2018,BoylanKolchin2008}, indicating a possible merger in $\gtrsim 400$ Myr.

Further evidence of an ongoing interaction between the main source and satellite is provided by recent imaging of the source by \acro{JWST} (see \autoref{fig:jwst}). Here both a bridge feature between the two systems and an approximately 20 kpc tidal tail streaming away from the satellite galaxy are apparent, confirming that the \jiooo pair are interacting.

Additionally the satellite's position relative to \jiooo is approximately perpendicular to the axis of rotation of \jiooo (see the moment-1 map in \autoref{fig:images}, bottom), which could be an indication that the companion was accreted from the same strand of the Cosmic Web that formed the main galaxy.

\subsection{Neutral Gas Cloud Parameters: cloud radius and numbers, volume, and area filling factors}
\label{sec:new_toy}

Molecular clouds are not singular units. They are complex entities fractured by the effects of star formation including the turbulent outflows from young stellar objects and winds and dissociating far-UV flux from high mass stars. It is challenging to directly discern the physical properties of molecular cloud distributions such as cloud size and filling factor in galaxies both near and far. This is because the common tracers of molecular clouds are the low-excitation CO lines in the millimeter wave regime where spatial resolution is modest and because the CO lines are optically thick so that complex internal structures within cloud complexes are well masked. Initial CO surveys of the Galaxy suggested size-scales of order $4\dashbetween40$ pc \citep{Solomon1987}, but later observations in the much more optically thin $^{13}$CO lines have shown that typical cloud sizes are much smaller, of the order 2 to 6 pc \citep{Heyer2009}. In external galaxies systematic surveys do not reach spatial resolutions below 10~pc, even for our neighbor *M31, where Submillimeter Array (\acro{SMA}) observations in CO(2-1) only reach spatial resolutions of 14 pc \citep{Lada2024}. High spatial resolution is much more difficult to obtain for high redshift systems where even for lensed galaxies like the ``Cosmic Snake'' the best linear resolution obtained is 30~pc \citep{DessaugesZavadsky2019}.

Fortunately, much as for stars, where their diameters are measured through a combination of physics (the radiation laws) and observations (parallax), the size of molecular clouds can be derived through a combination of astrophysics (\PDR models) and observations (\cii and CO line, and far-*IR continuum intensities).  In the following, we reintroduce this astrophysically based method that was
first demonstrated by \cite{Wolfire1990}.

In \autoref{sec:pdr} we used the observed \PDR line fluxes and far-\acro{IR} continuum flux and their ratios within a \PDR paradigm to derive the overall properties of the emitting medium including gas density $n_{H}$ and \FUV flux $G$ impinging on the emitting clouds (\autoref{fig:oiciifir}). Comparing \PDR parameters to the observed \cii line intensity we can derive important cloud parameters including cloud size, numbers of clouds and the volume and area filling factors within a toy model that assumes spherical clouds externally heated by the stellar \FUV radiation field. The details of this model are given in \autoref{sec:cloudsize}.

The key to our model is that we assume uniform density spherical clouds that are illuminated externally by a stellar \FUV radiation field of strength $G$.  The \cii line arises from the molecular cloud surfaces and the depth of the emission region is given by the \PDR models, so the cloud surface area and photodissociated skin depth $\Delta r$ are determined. The CO line intensity traces the molecular cloud mass which is proportional to the cloud volume.  Under the assumption that the \PDR has the same gas density as the molecular cloud, the line fluxes can then be used to derive the cloud size in a \PDR paradigm. From our \PDR models above we derived $n_{H}= 6 \times 10^{3}$ cm$^{-3}$  and $G = 5 \times 10^{3}\; G_{0}$. We adopt $M_m = 4.5\times10^{10}$ for the total molecular gas mass in \jiooo derived by \cite{Aravena2008} using a value for the observed CO intensity to molecular gas mass conversion factor $\alpha_{CO}=0.8 \tx{ M}_\odot$~(K km s$^{-1}$ pc$^2$ )$^{-1}$. Using our toy model we derive a beam filling factor $\phi_{A} = 4.2$, a cloud radius, $r_{CO}=3.4$ pc $\times \aco$, and a \cii line emitting shell depth of $\Delta r\sim 4.0\times 10^{21}$ cm$^{-2}/n_{H} =0.22$ pc.  There are $\sim 10^{6}$ clouds in an emission region of volume $2.3\times10^{11}$~pc$^{3}$, so the
cloud volume filling factor is $\phi_{vol}= 0.077\%$. The total mass of each individual cloud, including the molecular core and the photodissociated surface is $5.2\times 10^{4}\tx{ M}_\odot \times \aco^3$.

The cloud radius including the \PDR surface is 3.5 pc which is somewhat smaller than those inferred through CO measurements of nearby galaxies, but of the same order as those discerned in $^{13}$CO surveys. This is likely because the
\cii line arises from cloud surfaces so that \PDR-derived estimates are sensitive to the internal structures of molecular cloud complexes. Smaller scale structures would be expected for clouds turbulently stirred up and fractionated by recent intense episodes of star formation. It is also important to note that the cloud radius estimate from this model scales linearly with $M_{m}$, total molecular gas mass, which includes an assumption for the value of $\alpha_{CO}$.

\begin{table}
\centering
\caption{Observational parameters, parameters derived from the \PDR model
in \autoref{sec:pdr}, and assumptions used for deriving the filling factor in \jiooo. }
\label{tab:toyparams}
\begin{tabular}{l|l}
Metric                            & Value   \\
\hline
$z$                               & 1.8275 \\
$D_L$                             & 13.9 Gpc\\
{\cii} line flux                  & $37.7 \pm 3.3$ Jy km s$^{-1}$ \\
{\cii} luminosity                 & $(5.1\pm 0.4)\times10^{9}$ \lsun\\
{*FIR } luminosity                & $(7.2 \pm 0.3)\times10^{12}$ \lsun\\
{*SFR }                           & $1640 \pm 80$ \msun yr\inv\\
$n$                               & $6\cdot10^3$ cm$^{-3}\;\pm0.3$ dex\\
$G$                               & $5\cdot10^3\tx{ }G_0\;\pm0.3$ dex\\
$I_\textrm{CII}$                  & $10^{-3}\textrm{ ergs s}^{-1}\textrm{ cm}^{-2}\textrm{ sr}^{-1}$ \\
$R$                               & $0\farcs456 = 3.85$ kpc\\
$M_m$                             & $4.5\cdot10^{10}\textrm{ M}_\odot$\\
$\Omega$                          & $1.3\cdot10^{-11}$ sr $= 0.57$ arcsec$^2$\\

\end{tabular}
\end{table}

\subsection{Detailed Analysis of Masses}
With the data discussed thus far, we can constrain \sdssj's dynamical mass, stellar mass, and dust mass, and the satellite's stellar mass. In this section we report these values.

First, we calculate the dynamical mass using the results of our disk modeling in \autoref{sec:disk}. In general, the mass inside a certain radius is related to the rotational velocity by
$ M(r) = \frac{rV^2}{G}$. Substituting in our values, and assuming a flat rotation curve as found by the disk modeling (see \autoref{sec:modelresults}), we find
$M(r) = r\cdot3.0\cdot10^{10} \textrm{ \msun kpc}^{-1}$. Our disk model finds that the *FWHM of \cii brightness is $0\farcs45 \pm 0\farcs04$ or $3.8 \pm 0.3$ kpc. Using this as a radius, we estimate a dynamical mass of $\left(1.14 \pm 0.10\right)\cdot10^{11}$~\msun. This includes inclination effects. J*WST imaging in \autoref{fig:jwst} supports a much more extended source, with radius closer to $1\farcs0$. If we assume that the flat rotation curve extends to the J*WST radius of 8.4 kpc, the dynamical mass would be $\sim2.5\cdot10^{11}$ \msun.

This is compatible with stellar mass estimates from \cite{Liu2019}, who finds stellar mass $=7.9\cdot10^{10}$ \msun for \sdssj and $7.4\cdot10^{10}$ \msun for the satellite. Given the similarity of stellar masses between the satellite and \jiooo, and \cite{Liu2019}'s estimate that the *SFR in the satellite is a tenth that of the main source, we can surmise that the stellar population in the satellite is much older than that of \jiooo.

Ongoing intense star formation and the potential merger would add to this already massive ($\sim10^{11}$ \msun) stellar inventory, supporting an argument that \jiooo is evolving to be among the most massive galaxies seen today.

In \autoref{sec:sed} we used our Far-IR SED to constrain the dust mass in SDSSJ1000. We found a dust mass of $\left(1.4\pm0.1\right)\cdot10^9$ M$_\odot$. This is relatively consistent with previous dust mass estimates such as those presented by \cite{Aravena2008}.

\section{Summary and Outlook}
Within this work, we have investigated the star formation properties of \sdssj, an optically selected quasar at redshift 1.8275 with a large molecular gas reservoir that feeds an obscured starburst.  \jiooo was one of the first galaxies at Cosmic Noon to be detected in the 158 \micron \cii line, which inspired followup measurements of the 63 \micron \oi line with Herschel/*PACS and imaging of the \cii line at better than $0\farcs2$ spatial resolution with \ALMA.  Through comparison of these data with archival data we find:
\begin{itemize}

\item The \cii line emission is spatially resolved. The \cii velocity field is consistent with a dynamically cold rotating disk,  presented to us at an inclination of $\sim 36^{\circ}$ with a diameter $\geq 3.8$ kpc and maximum rotation velocity of 350 km/sec. Within this radius, the enclosed mass is dominated by that of the stellar population. The \cii emitting disk is enveloped by spiral arms visible in the rest-frame optical.
\item By comparing our \cii results with Herschel \oi observations and archival rest frame \FIR continuum observations, we show that the \cii emission is dominated by \PDRs with disk-averaged \FUV intensity $\sim5000$ G$_{0}$ and gas density $\sim6000$~cm$^{-3}$.
\item  Making use of the spatially resolved 158 \micron rest frame continuum as a proxy for the \FIR luminosity, we find that the \cii line to \FIR continuum luminosity ratio falls off with \FUV field strength in a manner consistent with predictions of \PDR models.  Therefore this ratio is a good proxy for the strength of the local \FUV fields, or star formation surface density as originally proposed in \cite{Wolfire1990} and utilized in the \cite{Stacey1991} survey.
\item We use a simple model based on our \PDR modeling results to derive typical cloud radii $r_{co}=3.4$ pc $\times \aco$ and mass $4.9\times10^4$ \msun $\times \aco^3$ in J1000 and both are smaller than values typical for nearby galaxies. This is likely the result of the intense star formation disrupting their natal cloud complexes.
\item We report the spectroscopic confirmation of a satellite galaxy within a projected distance of $\sim 16$ kpc of \jiooo. J*WST \acro{NIRC}am imaging reveals both a bridge of material between the two sources and a tidal tail feature extension for about 20~kpc roughly the opposite direction of the connecting bridge. We suggest that the system interaction axis may reflect the underlying dark matter distribution of the system and provide evidence of interaction.

\end{itemize}

These rare \ALMA Band 9 observations of a high-$z$ galaxy, consisting of only 2 hrs of telescope time, demonstrate capabilities of \ALMA that are seldom achieved.

\section{Acknowledgments}
CR thanks the National Research Council for their generous support through grant \texttt{506872.B3587}. This work was supported by the National Science Foundation through the following grants: \texttt{NRAO SOS 1519126, AST-1716229, AST-1910107, CAREER-1847892} and \texttt{AST-2009767}, and N\textsc{asa}/U*SRA S*OFIA grants \texttt{09-0185} and \texttt{07-0209}.

Herschel is an E*SA space observatory with science instruments provided by European-led Principal Investigator consortia and with important participation from N*ASA.

This paper makes use of the following \ALMA data: \texttt{ADS/JAO.ALMA \#2015.1.01362.S, \#2016.1.00463.S, \#2021.1.00246.S}, and \texttt{\#2021.1.00705.S}. \ALMA is a partnership of E*SO (representing its member states), N*SF (U\acold{SA}) and N*INS (Japan), together with N*RC (Canada), M*OST and A*SIAA (Taiwan), and K*ASI (Republic of Korea), in cooperation with the Republic of Chile. The Joint \ALMA Observatory is operated by *ESO, A\acold{UI}/N\acold{RAO} and N*AOJ. The National Radio Astronomy Observatory is a facility of the National Science Foundation operated under cooperative agreement by Associated Universities, Inc.

This work is based in part on observations made with the N\acold{ASA}/E\acold{SA}/C*SA James Webb Space Telescope. The data were obtained from the Mikulski Archive for Space Telescopes at the Space Telescope Science Institute, which is operated by the Association of Universities for Research in Astronomy, Inc., under N*ASA contract \texttt{NAS 5-03127} for J*WST. These observations are associated with program \texttt{\#1727}. The specific observations analyzed can be accessed via \dataset[doi: 10.17909/8sg7-7w28]{https://doi.org/10.17909/8sg7-7w28}.

This research made use of N\textsc{asa}'s Astrophysics Data System Bibliographic Services.

We thank the anonymous referee for their helpful comments which substantially improved this publication.

\shorthandoff
\bibliography{master}

\begin{thebibliography}{}
\expandafter\ifx\csname natexlab\endcsname\relax\def\natexlab#1{#1}\fi
\providecommand{\url}[1]{\href{#1}{#1}}
\providecommand{\dodoi}[1]{doi:~\href{http://doi.org/#1}{\nolinkurl{#1}}}
\providecommand{\doeprint}[1]{\href{http://ascl.net/#1}{\nolinkurl{http://ascl.net/#1}}}
\providecommand{\doarXiv}[1]{\href{https://arxiv.org/abs/#1}{\nolinkurl{https://arxiv.org/abs/#1}}}

\bibitem[{{Aravena} {et~al.}(2008){Aravena}, {Bertoldi}, {Schinnerer}, {Weiss}, {Jahnke}, {Carilli}, {Frayer}, {Henkel}, {Brusa}, {Menten}, {Salvato}, \& {Smolcic}}]{Aravena2008}
{Aravena}, M., {Bertoldi}, F., {Schinnerer}, E., {et~al.} 2008, \aap, 491, 173, \dodoi{10.1051/0004-6361:200810628}

\bibitem[{{Atek} {et~al.}(2023){Atek}, {Shuntov}, {Furtak}, {Richard}, {Kneib}, {Mahler}, {Zitrin}, {McCracken}, {Charlot}, {Chevallard}, \& {Chemerynska}}]{Atek2023}
{Atek}, H., {Shuntov}, M., {Furtak}, L.~J., {et~al.} 2023, \mnras, 519, 1201, \dodoi{10.1093/mnras/stac3144}

\bibitem[{{Bertoldi} {et~al.}(2007){Bertoldi}, {Carilli}, {Aravena}, {Schinnerer}, {Voss}, {Smolcic}, {Jahnke}, {Scoville}, {Blain}, {Menten}, {Lutz}, {Brusa}, {Taniguchi}, {Capak}, {Mobasher}, {Lilly}, {Thompson}, {Aussel}, {Kreysa}, {Hasinger}, {Aguirre}, {Schlaerth}, \& {Koekemoer}}]{Bertoldi2007}
{Bertoldi}, F., {Carilli}, C., {Aravena}, M., {et~al.} 2007, \apjs, 172, 132, \dodoi{10.1086/520511}

\bibitem[{{Bouwens} {et~al.}(2022){Bouwens}, {Smit}, {Schouws}, {Stefanon}, {Bowler}, {Endsley}, {Gonzalez}, {Inami}, {Stark}, {Oesch}, {Hodge}, {Aravena}, {da Cunha}, {Dayal}, {de Looze}, {Ferrara}, {Fudamoto}, {Graziani}, {Li}, {Nanayakkara}, {Pallottini}, {Schneider}, {Sommovigo}, {Topping}, {van der Werf}, {Algera}, {Barrufet}, {Hygate}, {Labb{\'e}}, {Riechers}, \& {Witstok}}]{Bouwens2022}
{Bouwens}, R.~J., {Smit}, R., {Schouws}, S., {et~al.} 2022, \apj, 931, 160, \dodoi{10.3847/1538-4357/ac5a4a}

\bibitem[{{Boylan-Kolchin} {et~al.}(2008){Boylan-Kolchin}, {Ma}, \& {Quataert}}]{BoylanKolchin2008}
{Boylan-Kolchin}, M., {Ma}, C.-P., \& {Quataert}, E. 2008, \mnras, 383, 93, \dodoi{10.1111/j.1365-2966.2007.12530.x}

\bibitem[{{Brauher} {et~al.}(2008){Brauher}, {Dale}, \& {Helou}}]{Brauher2008}
{Brauher}, J.~R., {Dale}, D.~A., \& {Helou}, G. 2008, \apjs, 178, 280, \dodoi{10.1086/590249}

\bibitem[{{Brisbin} {et~al.}(2015){Brisbin}, {Ferkinhoff}, {Nikola}, {Parshley}, {Stacey}, {Spoon}, {Hailey-Dunsheath}, \& {Verma}}]{Brisbin2015}
{Brisbin}, D., {Ferkinhoff}, C., {Nikola}, T., {et~al.} 2015, \apj, 799, 13, \dodoi{10.1088/0004-637X/799/1/13}

\bibitem[{{Buchner} {et~al.}(2014){Buchner}, {Georgakakis}, {Nandra}, {Hsu}, {Rangel}, {Brightman}, {Merloni}, {Salvato}, {Donley}, \& {Kocevski}}]{Buchner2014}
{Buchner}, J., {Georgakakis}, A., {Nandra}, K., {et~al.} 2014, \aap, 564, A125, \dodoi{10.1051/0004-6361/201322971}

\bibitem[{{Carilli} \& {Walter}(2013)}]{Carilli2013}
{Carilli}, C.~L., \& {Walter}, F. 2013, \araa, 51, 105, \dodoi{10.1146/annurev-astro-082812-140953}

\bibitem[{{Casey} {et~al.}(2023){Casey}, {Kartaltepe}, {Drakos}, {Franco}, {Harish}, {Paquereau}, {Ilbert}, {Rose}, {Cox}, {Nightingale}, {Robertson}, {Silverman}, {Koekemoer}, {Massey}, {McCracken}, {Rhodes}, {Akins}, {Allen}, {Amvrosiadis}, {Arango-Toro}, {Bagley}, {Bongiorno}, {Capak}, {Champagne}, {Chartab}, {Ch{\'a}vez Ortiz}, {Chworowsky}, {Cooke}, {Cooper}, {Darvish}, {Ding}, {Faisst}, {Finkelstein}, {Fujimoto}, {Gentile}, {Gillman}, {Gould}, {Gozaliasl}, {Hayward}, {He}, {Hemmati}, {Hirschmann}, {Jahnke}, {Jin}, {Khostovan}, {Kokorev}, {Lambrides}, {Laigle}, {Larson}, {Leung}, {Liu}, {Liaudat}, {Long}, {Magdis}, {Mahler}, {Mainieri}, {Manning}, {Maraston}, {Martin}, {McCleary}, {McKinney}, {McPartland}, {Mobasher}, {Pattnaik}, {Renzini}, {Rich}, {Sanders}, {Sattari}, {Scognamiglio}, {Scoville}, {Sheth}, {Shuntov}, {Sparre}, {Suzuki}, {Talia}, {Toft}, {Trakhtenbrot}, {Urry}, {Valentino}, {Vanderhoof}, {Vardoulaki}, {Weaver}, {Whitaker}, {Wilkins}, {Yang}, \&
  {Zavala}}]{Casey2023}
{Casey}, C.~M., {Kartaltepe}, J.~S., {Drakos}, N.~E., {et~al.} 2023, \apj, 954, 31, \dodoi{10.3847/1538-4357/acc2bc}

\bibitem[{{Crawford} {et~al.}(1985){Crawford}, {Genzel}, {Townes}, \& {Watson}}]{Crawford1985}
{Crawford}, M.~K., {Genzel}, R., {Townes}, C.~H., \& {Watson}, D.~M. 1985, \apj, 291, 755, \dodoi{10.1086/163113}

\bibitem[{{Curtis-Lake} {et~al.}(2023){Curtis-Lake}, {Carniani}, {Cameron}, {Charlot}, {Jakobsen}, {Maiolino}, {Bunker}, {Witstok}, {Smit}, {Chevallard}, {Willott}, {Ferruit}, {Arribas}, {Bonaventura}, {Curti}, {D'Eugenio}, {Franx}, {Giardino}, {Looser}, {L{\"u}tzgendorf}, {Maseda}, {Rawle}, {Rix}, {Rodr{\'\i}guez del Pino}, {{\"U}bler}, {Sirianni}, {Dressler}, {Egami}, {Eisenstein}, {Endsley}, {Hainline}, {Hausen}, {Johnson}, {Rieke}, {Robertson}, {Shivaei}, {Stark}, {Tacchella}, {Williams}, {Willmer}, {Bhatawdekar}, {Bowler}, {Boyett}, {Chen}, {de Graaff}, {Helton}, {Hviding}, {Jones}, {Kumari}, {Lyu}, {Nelson}, {Perna}, {Sandles}, {Saxena}, {Suess}, {Sun}, {Topping}, {Wallace}, \& {Whitler}}]{CurtisLake2023}
{Curtis-Lake}, E., {Carniani}, S., {Cameron}, A., {et~al.} 2023, Nature Astronomy, 7, 622, \dodoi{10.1038/s41550-023-01918-w}

\bibitem[{{Dale} \& {Helou}(2002)}]{Dale2002}
{Dale}, D.~A., \& {Helou}, G. 2002, \apj, 576, 159, \dodoi{10.1086/341632}

\bibitem[{{Davis} {et~al.}(2013){Davis}, {Alatalo}, {Bureau}, {Cappellari}, {Scott}, {Young}, {Blitz}, {Crocker}, {Bayet}, {Bois}, {Bournaud}, {Davies}, {de Zeeuw}, {Duc}, {Emsellem}, {Khochfar}, {Krajnovi{\'c}}, {Kuntschner}, {Lablanche}, {McDermid}, {Morganti}, {Naab}, {Oosterloo}, {Sarzi}, {Serra}, \& {Weijmans}}]{Davis2013}
{Davis}, T.~A., {Alatalo}, K., {Bureau}, M., {et~al.} 2013, \mnras, 429, 534, \dodoi{10.1093/mnras/sts353}

\bibitem[{{Decarli} {et~al.}(2018){Decarli}, {Walter}, {Venemans}, {Ba{\~n}ados}, {Bertoldi}, {Carilli}, {Fan}, {Farina}, {Mazzucchelli}, {Riechers}, {Rix}, {Strauss}, {Wang}, \& {Yang}}]{Decarli2018}
{Decarli}, R., {Walter}, F., {Venemans}, B.~P., {et~al.} 2018, \apj, 854, 97, \dodoi{10.3847/1538-4357/aaa5aa}

\bibitem[{{Dessauges-Zavadsky} {et~al.}(2019){Dessauges-Zavadsky}, {Richard}, {Combes}, {Schaerer}, {Rujopakarn}, {Mayer}, {Cava}, {Boone}, {Egami}, {Kneib}, {P{\'e}rez-Gonz{\'a}lez}, {Pfenniger}, {Rawle}, {Teyssier}, \& {van der Werf}}]{DessaugesZavadsky2019}
{Dessauges-Zavadsky}, M., {Richard}, J., {Combes}, F., {et~al.} 2019, Nature Astronomy, 3, 1115, \dodoi{10.1038/s41550-019-0874-0}

\bibitem[{{D{\'\i}az-Santos} {et~al.}(2013){D{\'\i}az-Santos}, {Armus}, {Charmandaris}, {Stierwalt}, {Murphy}, {Haan}, {Inami}, {Malhotra}, {Meijerink}, {Stacey}, {Petric}, {Evans}, {Veilleux}, {van der Werf}, {Lord}, {Lu}, {Howell}, {Appleton}, {Mazzarella}, {Surace}, {Xu}, {Schulz}, {Sanders}, {Bridge}, {Chan}, {Frayer}, {Iwasawa}, {Melbourne}, \& {Sturm}}]{DiazSantos2013}
{D{\'\i}az-Santos}, T., {Armus}, L., {Charmandaris}, V., {et~al.} 2013, \apj, 774, 68, \dodoi{10.1088/0004-637X/774/1/68}

\bibitem[{{D{\'{\i}}az-Santos} {et~al.}(2017){D{\'{\i}}az-Santos}, {Armus}, {Charmandaris}, {Lu}, {Stierwalt}, {Stacey}, {Malhotra}, {van der Werf}, {Howell}, {Privon}, {Mazzarella}, {Goldsmith}, {Murphy}, {Barcos-Mu{\~n}oz}, {Linden}, {Inami}, {Larson}, {Evans}, {Appleton}, {Iwasawa}, {Lord}, {Sanders}, \& {Surace}}]{DiazSantos2017}
{D{\'{\i}}az-Santos}, T., {Armus}, L., {Charmandaris}, V., {et~al.} 2017, \apj, 846, 32, \dodoi{10.3847/1538-4357/aa81d7}

\bibitem[{{Ferkinhoff}(2015)}]{Ferkinhoff2015a}
{Ferkinhoff}, C. 2015, in American Astronomical Society Meeting Abstracts, Vol. 225, American Astronomical Society Meeting Abstracts \#225, 141.19

\bibitem[{{Ferkinhoff} {et~al.}(2014){Ferkinhoff}, {Brisbin}, {Parshley}, {Nikola}, {Stacey}, {Schoenwald}, {Higdon}, {Higdon}, {Verma}, {Riechers}, {Hailey-Dunsheath}, {Menten}, {G{\"u}sten}, {Wei{\ss}}, {Irwin}, {Cho}, {Niemack}, {Halpern}, {Amiri}, {Hasselfield}, {Wiebe}, {Ade}, \& {Tucker}}]{Ferkinhoff2014a}
{Ferkinhoff}, C., {Brisbin}, D., {Parshley}, S., {et~al.} 2014, \apj, 780, 142, \dodoi{10.1088/0004-637X/780/2/142}

\bibitem[{{Fern{\'a}ndez Aranda} {et~al.}(2024){Fern{\'a}ndez Aranda}, {D{\'\i}az Santos}, {Hatziminaoglou}, {Assef}, {Aravena}, {Eisenhardt}, {Ferkinhoff}, {Pensabene}, {Nikola}, {Andreani}, {Vishwas}, {Stacey}, {Decarli}, {Blain}, {Brisbin}, {Charmandaris}, {Jun}, {Li}, {Liao}, {Martin}, {Stern}, {Tsai}, {Wu}, \& {Zewdie}}]{FernandezAranda2024}
{Fern{\'a}ndez Aranda}, R., {D{\'\i}az Santos}, T., {Hatziminaoglou}, E., {et~al.} 2024, \aap, 682, A166, \dodoi{10.1051/0004-6361/202347869}

\bibitem[{{Feroz} {et~al.}(2009){Feroz}, {Hobson}, \& {Bridges}}]{Feroz2009}
{Feroz}, F., {Hobson}, M.~P., \& {Bridges}, M. 2009, \mnras, 398, 1601, \dodoi{10.1111/j.1365-2966.2009.14548.x}

\bibitem[{{Frias Castillo} {et~al.}(2024){Frias Castillo}, {Rybak}, {Hodge}, {van der Werf}, {Abbo}, {Ballieux}, {Ward}, {Harrison}, {Calistro Rivera}, {McKean}, \& {Stacey}}]{FriasCastillo2024}
{Frias Castillo}, M., {Rybak}, M., {Hodge}, J., {et~al.} 2024, \aap, 683, A211, \dodoi{10.1051/0004-6361/202347596}

\bibitem[{{Gullberg} {et~al.}(2015){Gullberg}, {De Breuck}, {Vieira}, {Wei{\ss}}, {Aguirre}, {Aravena}, {B{\'e}thermin}, {Bradford}, {Bothwell}, {Carlstrom}, {Chapman}, {Fassnacht}, {Gonzalez}, {Greve}, {Hezaveh}, {Holzapfel}, {Husband}, {Ma}, {Malkan}, {Marrone}, {Menten}, {Murphy}, {Reichardt}, {Spilker}, {Stark}, {Strandet}, \& {Welikala}}]{Gullberg2015}
{Gullberg}, B., {De Breuck}, C., {Vieira}, J.~D., {et~al.} 2015, \mnras, 449, 2883, \dodoi{10.1093/mnras/stv372}

\bibitem[{{Gullberg} {et~al.}(2018){Gullberg}, {Swinbank}, {Smail}, {Biggs}, {Bertoldi}, {De Breuck}, {Chapman}, {Chen}, {Cooke}, {Coppin}, {Cox}, {Dannerbauer}, {Dunlop}, {Edge}, {Farrah}, {Geach}, {Greve}, {Hodge}, {Ibar}, {Ivison}, {Karim}, {Schinnerer}, {Scott}, {Simpson}, {Stach}, {Thomson}, {van der Werf}, {Walter}, {Wardlow}, \& {Weiss}}]{Gullberg2018}
{Gullberg}, B., {Swinbank}, A.~M., {Smail}, I., {et~al.} 2018, \apj, 859, 12, \dodoi{10.3847/1538-4357/aabe8c}

\bibitem[{{Hailey-Dunsheath} {et~al.}(2010){Hailey-Dunsheath}, {Nikola}, {Stacey}, {Oberst}, {Parshley}, {Benford}, {Staguhn}, \& {Tucker}}]{HaileyDunsheath2010}
{Hailey-Dunsheath}, S., {Nikola}, T., {Stacey}, G.~J., {et~al.} 2010, \apjl, 714, L162, \dodoi{10.1088/2041-8205/714/1/L162}

\bibitem[{{Herrera-Camus} {et~al.}(2018){Herrera-Camus}, {Sturm}, {Graci{\'a}-Carpio}, {Lutz}, {Contursi}, {Veilleux}, {Fischer}, {Gonz{\'a}lez-Alfonso}, {Poglitsch}, {Tacconi}, {Genzel}, {Maiolino}, {Sternberg}, {Davies}, \& {Verma}}]{HerreraCamus2018b}
{Herrera-Camus}, R., {Sturm}, E., {Graci{\'a}-Carpio}, J., {et~al.} 2018, \apj, 861, 95, \dodoi{10.3847/1538-4357/aac0f9}

\bibitem[{{Heyer} {et~al.}(2009){Heyer}, {Krawczyk}, {Duval}, \& {Jackson}}]{Heyer2009}
{Heyer}, M., {Krawczyk}, C., {Duval}, J., \& {Jackson}, J.~M. 2009, \apj, 699, 1092, \dodoi{10.1088/0004-637X/699/2/1092}

\bibitem[{{Houck} {et~al.}(1985){Houck}, {Schneider}, {Danielson}, {Beichman}, {Lonsdale}, {Neugebauer}, \& {Soifer}}]{Houck1985}
{Houck}, J.~R., {Schneider}, D.~P., {Danielson}, G.~E., {et~al.} 1985, \apjl, 290, L5, \dodoi{10.1086/184431}

\bibitem[{{Ishii} {et~al.}(2024){Ishii}, {Hashimoto}, {Ferkinhoff}, {Rybak}, {Inoue}, {Michiyama}, {Donevski}, {Fujimoto}, {Salak}, {Kuno}, {Matsuo}, {Mawatari}, {Tamura}, {Izumi}, {Nagao}, {Nakazato}, {Osone}, {Sugahara}, {Usui}, {Wakasugi}, {Yajima}, {Bakx}, {Fudamoto}, {Meyer}, {Walter}, \& {Yoshida}}]{Ishii2024}
{Ishii}, N., {Hashimoto}, T., {Ferkinhoff}, C., {et~al.} 2024, arXiv e-prints, arXiv:2408.09944, \dodoi{10.48550/arXiv.2408.09944}

\bibitem[{{Kaufman} {et~al.}(2006){Kaufman}, {Wolfire}, \& {Hollenbach}}]{Kaufman2006}
{Kaufman}, M.~J., {Wolfire}, M.~G., \& {Hollenbach}, D.~J. 2006, \apj, 644, 283, \dodoi{10.1086/503596}

\bibitem[{{Kaufman} {et~al.}(1999){Kaufman}, {Wolfire}, {Hollenbach}, \& {Luhman}}]{Kaufman1999}
{Kaufman}, M.~J., {Wolfire}, M.~G., {Hollenbach}, D.~J., \& {Luhman}, M.~L. 1999, \apj, 527, 795, \dodoi{10.1086/308102}

\bibitem[{{Kennicutt}(1998)}]{Kennicutt1998}
{Kennicutt}, Jr., R.~C. 1998, \araa, 36, 189, \dodoi{10.1146/annurev.astro.36.1.189}

\bibitem[{{Lada} {et~al.}(2024){Lada}, {Forbrich}, {Petitpas}, \& {Viaene}}]{Lada2024}
{Lada}, C.~J., {Forbrich}, J., {Petitpas}, G., \& {Viaene}, S. 2024, \apj, 966, 193, \dodoi{10.3847/1538-4357/ad38bf}

\bibitem[{{Lamarche} {et~al.}(2018){Lamarche}, {Verma}, {Vishwas}, {Stacey}, {Brisbin}, {Ferkinhoff}, {Nikola}, {Higdon}, {Higdon}, \& {Tecza}}]{Lamarche2018}
{Lamarche}, C., {Verma}, A., {Vishwas}, A., {et~al.} 2018, \apj, 867, 140, \dodoi{10.3847/1538-4357/aae394}

\bibitem[{{Litke} {et~al.}(2019){Litke}, {Marrone}, {Spilker}, {Aravena}, {B{\'e}thermin}, {Chapman}, {Chen}, {de Breuck}, {Dong}, {Gonzalez}, {Greve}, {Hayward}, {Hezaveh}, {Jarugula}, {Ma}, {Morningstar}, {Narayanan}, {Phadke}, {Reuter}, {Vieira}, \& {Weiss}}]{Litke2019}
{Litke}, K.~C., {Marrone}, D.~P., {Spilker}, J.~S., {et~al.} 2019, \apj, 870, 80, \dodoi{10.3847/1538-4357/aaf057}

\bibitem[{{Liu} {et~al.}(2019){Liu}, {Lang}, {Magnelli}, {Schinnerer}, {Leslie}, {Fudamoto}, {Bondi}, {Groves}, {Jim{\'e}nez-Andrade}, {Harrington}, {Karim}, {Oesch}, {Sargent}, {Vardoulaki}, {B{\v{a}}descu}, {Moser}, {Bertoldi}, {Battisti}, {da Cunha}, {Zavala}, {Vaccari}, {Davidzon}, {Riechers}, \& {Aravena}}]{Liu2019}
{Liu}, D., {Lang}, P., {Magnelli}, B., {et~al.} 2019, \apjs, 244, 40, \dodoi{10.3847/1538-4365/ab42da}

\bibitem[{{Luhman} {et~al.}(1998){Luhman}, {Satyapal}, {Fischer}, {Wolfire}, {Cox}, {Lord}, {Smith}, {Stacey}, \& {Unger}}]{Luhman1998}
{Luhman}, M.~L., {Satyapal}, S., {Fischer}, J., {et~al.} 1998, \apjl, 504, L11, \dodoi{10.1086/311562}

\bibitem[{{Madau} \& {Dickinson}(2014)}]{Madau2014}
{Madau}, P., \& {Dickinson}, M. 2014, \araa, 52, 415, \dodoi{10.1146/annurev-astro-081811-125615}

\bibitem[{{Malhotra} {et~al.}(2001){Malhotra}, {Kaufman}, {Hollenbach}, {Helou}, {Rubin}, {Brauher}, {Dale}, {Lu}, {Lord}, {Stacey}, {Contursi}, {Hunter}, \& {Dinerstein}}]{Malhotra2001}
{Malhotra}, S., {Kaufman}, M.~J., {Hollenbach}, D., {et~al.} 2001, \apj, 561, 766, \dodoi{10.1086/323046}

\bibitem[{{Marton} {et~al.}(2017){Marton}, {Calzoletti}, {Perez Garcia}, {Kiss}, {Paladini}, {Altieri}, {Sanchez Portal}, {Kidger}, \& {the Herschel Point Source Catalogue Working Group}}]{Marton2017}
{Marton}, G., {Calzoletti}, L., {Perez Garcia}, A.~M., {et~al.} 2017, arXiv e-prints, arXiv:1705.05693.
\newblock \doarXiv{1705.05693}

\bibitem[{{Mashian} {et~al.}(2015){Mashian}, {Sturm}, {Sternberg}, {Janssen}, {Hailey-Dunsheath}, {Fischer}, {Contursi}, {Gonz{\'a}lez-Alfonso}, {Graci{\'a}-Carpio}, {Poglitsch}, {Veilleux}, {Davies}, {Genzel}, {Lutz}, {Tacconi}, {Verma}, {Wei{\ss}}, {Polisensky}, \& {Nikola}}]{Mashian2015}
{Mashian}, N., {Sturm}, E., {Sternberg}, A., {et~al.} 2015, \apj, 802, 81, \dodoi{10.1088/0004-637X/802/2/81}

\bibitem[{{McKinney} {et~al.}(2020){McKinney}, {Pope}, {Armus}, {Chary}, {D{\'\i}az-Santos}, {Dickinson}, \& {Kirkpatrick}}]{McKinney2020}
{McKinney}, J., {Pope}, A., {Armus}, L., {et~al.} 2020, \apj, 892, 119, \dodoi{10.3847/1538-4357/ab77b9}

\bibitem[{{McMullin} {et~al.}(2007){McMullin}, {Waters}, {Schiebel}, {Young}, \& {Golap}}]{McMullin2007}
{McMullin}, J.~P., {Waters}, B., {Schiebel}, D., {Young}, W., \& {Golap}, K. 2007, in Astronomical Society of the Pacific Conference Series, Vol. 376, Astronomical Data Analysis Software and Systems XVI, ed. R.~A. {Shaw}, F.~{Hill}, \& D.~J. {Bell}, 127

\bibitem[{{Oberst} {et~al.}(2006){Oberst}, {Parshley}, {Stacey}, {Nikola}, {L{\"o}hr}, {Harnett}, {Tothill}, {Lane}, {Stark}, \& {Tucker}}]{Oberst2006}
{Oberst}, T.~E., {Parshley}, S.~C., {Stacey}, G.~J., {et~al.} 2006, \apjl, 652, L125, \dodoi{10.1086/510289}

\bibitem[{{Ott}(2010)}]{Ott2010}
{Ott}, S. 2010, in Astronomical Society of the Pacific Conference Series, Vol. 434, Astronomical Data Analysis Software and Systems XIX, ed. Y.~{Mizumoto}, K.~I. {Morita}, \& M.~{Ohishi}, 139, \dodoi{10.48550/arXiv.1011.1209}

\bibitem[{{Pavesi} {et~al.}(2018){Pavesi}, {Riechers}, {Sharon}, {Smol{\v{c}}i{\'c}}, {Faisst}, {Schinnerer}, {Carilli}, {Capak}, {Scoville}, \& {Stacey}}]{Pavesi2018}
{Pavesi}, R., {Riechers}, D.~A., {Sharon}, C.~E., {et~al.} 2018, \apj, 861, 43, \dodoi{10.3847/1538-4357/aac6b6}

\bibitem[{{Perna} {et~al.}(2018){Perna}, {Sargent}, {Brusa}, {Daddi}, {Feruglio}, {Cresci}, {Lanzuisi}, {Lusso}, {Comastri}, {Coogan}, {D'Amato}, {Gilli}, {Piconcelli}, \& {Vignali}}]{Perna2018}
{Perna}, M., {Sargent}, M.~T., {Brusa}, M., {et~al.} 2018, \aap, 619, A90, \dodoi{10.1051/0004-6361/201833040}

\bibitem[{{Pillepich} {et~al.}(2019){Pillepich}, {Nelson}, {Springel}, {Pakmor}, {Torrey}, {Weinberger}, {Vogelsberger}, {Marinacci}, {Genel}, {van der Wel}, \& {Hernquist}}]{Pillepich2019}
{Pillepich}, A., {Nelson}, D., {Springel}, V., {et~al.} 2019, \mnras, 490, 3196, \dodoi{10.1093/mnras/stz2338}

\bibitem[{{Poglitsch} {et~al.}(2010){Poglitsch}, {Waelkens}, {Geis}, {Feuchtgruber}, {Vandenbussche}, {Rodriguez}, {Krause}, {Renotte}, {van Hoof}, {Saraceno}, {Cepa}, {Kerschbaum}, {Agn{\`e}se}, {Ali}, {Altieri}, {Andreani}, {Augueres}, {Balog}, {Barl}, {Bauer}, {Belbachir}, {Benedettini}, {Billot}, {Boulade}, {Bischof}, {Blommaert}, {Callut}, {Cara}, {Cerulli}, {Cesarsky}, {Contursi}, {Creten}, {De Meester}, {Doublier}, {Doumayrou}, {Duband}, {Exter}, {Genzel}, {Gillis}, {Gr{\"o}zinger}, {Henning}, {Herreros}, {Huygen}, {Inguscio}, {Jakob}, {Jamar}, {Jean}, {de Jong}, {Katterloher}, {Kiss}, {Klaas}, {Lemke}, {Lutz}, {Madden}, {Marquet}, {Martignac}, {Mazy}, {Merken}, {Montfort}, {Morbidelli}, {M{\"u}ller}, {Nielbock}, {Okumura}, {Orfei}, {Ottensamer}, {Pezzuto}, {Popesso}, {Putzeys}, {Regibo}, {Reveret}, {Royer}, {Sauvage}, {Schreiber}, {Stegmaier}, {Schmitt}, {Schubert}, {Sturm}, {Thiel}, {Tofani}, {Vavrek}, {Wetzstein}, {Wieprecht}, \& {Wiezorrek}}]{Poglitsch2010}
{Poglitsch}, A., {Waelkens}, C., {Geis}, N., {et~al.} 2010, \aap, 518, L2, \dodoi{10.1051/0004-6361/201014535}

\bibitem[{{Pound} \& {Wolfire}(2008)}]{Pound2008}
{Pound}, M.~W., \& {Wolfire}, M.~G. 2008, in Astronomical Society of the Pacific Conference Series, Vol. 394, Astronomical Data Analysis Software and Systems XVII, ed. R.~W. {Argyle}, P.~S. {Bunclark}, \& J.~R. {Lewis}, 654

\bibitem[{{Pound} \& {Wolfire}(2011)}]{Pound2011}
{Pound}, M.~W., \& {Wolfire}, M.~G. 2011, PDRT: Photo Dissociation Region Toolbox, Astrophysics Source Code Library, record ascl:1102.022.
\newblock \doeprint{1102.022}

\bibitem[{{Pound} \& {Wolfire}(2023)}]{Pound2023}
---. 2023, \aj, 165, 25, \dodoi{10.3847/1538-3881/ac9b1f}

\bibitem[{{Rigopoulou} {et~al.}(2018){Rigopoulou}, {Pereira-Santaella}, {Magdis}, {Cooray}, {Farrah}, {Marques-Chaves}, {Perez-Fournon}, \& {Riechers}}]{Rigopoulou2018}
{Rigopoulou}, D., {Pereira-Santaella}, M., {Magdis}, G.~E., {et~al.} 2018, \mnras, 473, 20, \dodoi{10.1093/mnras/stx2311}

\bibitem[{{Rizzo} {et~al.}(2022){Rizzo}, {Kohandel}, {Pallottini}, {Zanella}, {Ferrara}, {Vallini}, \& {Toft}}]{Rizzo2022}
{Rizzo}, F., {Kohandel}, M., {Pallottini}, A., {et~al.} 2022, \aap, 667, A5, \dodoi{10.1051/0004-6361/202243582}

\bibitem[{{Rizzo} {et~al.}(2020){Rizzo}, {Vegetti}, {Powell}, {Fraternali}, {McKean}, {Stacey}, \& {White}}]{Rizzo2020}
{Rizzo}, F., {Vegetti}, S., {Powell}, D., {et~al.} 2020, \nat, 584, 201, \dodoi{10.1038/s41586-020-2572-6}

\bibitem[{{Rybak} {et~al.}(2020{\natexlab{a}}){Rybak}, {Hodge}, {Vegetti}, {van der Werf}, {Andreani}, {Graziani}, \& {McKean}}]{Rybak2020b}
{Rybak}, M., {Hodge}, J.~A., {Vegetti}, S., {et~al.} 2020{\natexlab{a}}, \mnras, 494, 5542, \dodoi{10.1093/mnras/staa879}

\bibitem[{{Rybak} {et~al.}(2020{\natexlab{b}}){Rybak}, {Zavala}, {Hodge}, {Casey}, \& {Werf}}]{Rybak2020a}
{Rybak}, M., {Zavala}, J.~A., {Hodge}, J.~A., {Casey}, C.~M., \& {Werf}, P. v.~d. 2020{\natexlab{b}}, \apjl, 889, L11, \dodoi{10.3847/2041-8213/ab63de}

\bibitem[{{Rybak} {et~al.}(2019){Rybak}, {Calistro Rivera}, {Hodge}, {Smail}, {Walter}, {van der Werf}, {da Cunha}, {Chen}, {Dannerbauer}, {Ivison}, {Karim}, {Simpson}, {Swinbank}, \& {Wardlow}}]{Rybak2019}
{Rybak}, M., {Calistro Rivera}, G., {Hodge}, J.~A., {et~al.} 2019, \apj, 876, 112, \dodoi{10.3847/1538-4357/ab0e0f}

\bibitem[{{Rybak} {et~al.}(2021){Rybak}, {da Cunha}, {Groves}, {Hodge}, {Aravena}, {Maseda}, {Boogaard}, {Berg}, {Charlot}, {Decarli}, {Erb}, {Nelson}, {Pacifici}, {Schmidt}, {Walter}, \& {van der Wel}}]{Rybak2021}
{Rybak}, M., {da Cunha}, E., {Groves}, B., {et~al.} 2021, \apj, 909, 130, \dodoi{10.3847/1538-4357/abd946}

\bibitem[{{Sanders} {et~al.}(1988){Sanders}, {Soifer}, {Elias}, {Madore}, {Matthews}, {Neugebauer}, \& {Scoville}}]{Sanders1988}
{Sanders}, D.~B., {Soifer}, B.~T., {Elias}, J.~H., {et~al.} 1988, \apj, 325, 74, \dodoi{10.1086/165983}

\bibitem[{{Sanders} {et~al.}(2007){Sanders}, {Salvato}, {Aussel}, {Ilbert}, {Scoville}, {Surace}, {Frayer}, {Sheth}, {Helou}, {Brooke}, {Bhattacharya}, {Yan}, {Kartaltepe}, {Barnes}, {Blain}, {Calzetti}, {Capak}, {Carilli}, {Carollo}, {Comastri}, {Daddi}, {Ellis}, {Elvis}, {Fall}, {Franceschini}, {Giavalisco}, {Hasinger}, {Impey}, {Koekemoer}, {Le F{\`e}vre}, {Lilly}, {Liu}, {McCracken}, {Mobasher}, {Renzini}, {Rich}, {Schinnerer}, {Shopbell}, {Taniguchi}, {Thompson}, {Urry}, \& {Williams}}]{Sanders2007}
{Sanders}, D.~B., {Salvato}, M., {Aussel}, H., {et~al.} 2007, \apjs, 172, 86, \dodoi{10.1086/517885}

\bibitem[{{Schaerer} {et~al.}(2015){Schaerer}, {Boone}, {Jones}, {Dessauges-Zavadsky}, {Sklias}, {Zamojski}, {Cava}, {Richard}, {Ellis}, {Rawle}, {Egami}, \& {Combes}}]{Schaerer2015}
{Schaerer}, D., {Boone}, F., {Jones}, T., {et~al.} 2015, \aap, 576, L2, \dodoi{10.1051/0004-6361/201425542}

\bibitem[{{Schaerer} {et~al.}(2020){Schaerer}, {Ginolfi}, {B{\'e}thermin}, {Fudamoto}, {Oesch}, {Le F{\`e}vre}, {Faisst}, {Capak}, {Cassata}, {Silverman}, {Yan}, {Jones}, {Amorin}, {Bardelli}, {Boquien}, {Cimatti}, {Dessauges-Zavadsky}, {Giavalisco}, {Hathi}, {Fujimoto}, {Ibar}, {Koekemoer}, {Lagache}, {Lemaux}, {Loiacono}, {Maiolino}, {Narayanan}, {Morselli}, {M{\'e}ndez-Hern{\`a}ndez}, {Pozzi}, {Riechers}, {Talia}, {Toft}, {Vallini}, {Vergani}, {Zamorani}, \& {Zucca}}]{Schaerer2020}
{Schaerer}, D., {Ginolfi}, M., {B{\'e}thermin}, M., {et~al.} 2020, \aap, 643, A3, \dodoi{10.1051/0004-6361/202037617}

\bibitem[{{Schulz} {et~al.}(2017){Schulz}, {Marton}, {Valtchanov}, {Mar{\'\i}a P{\'e}rez Garc{\'\i}a}, {Pint{\'e}r}, {Appleton}, {Kiss}, {Lim}, {Lu}, {Papageorgiou}, {Pearson}, {Rector}, {S{\'a}nchez Portal}, {Shupe}, {T{\'o}th}, {Van Dyk}, {Varga-Vereb{\'e}lyi}, \& {Xu}}]{Schulz2017}
{Schulz}, B., {Marton}, G., {Valtchanov}, I., {et~al.} 2017, arXiv e-prints, arXiv:1706.00448.
\newblock \doarXiv{1706.00448}

\bibitem[{{Scoville} {et~al.}(2007){Scoville}, {Aussel}, {Brusa}, {Capak}, {Carollo}, {Elvis}, {Giavalisco}, {Guzzo}, {Hasinger}, {Impey}, {Kneib}, {LeFevre}, {Lilly}, {Mobasher}, {Renzini}, {Rich}, {Sanders}, {Schinnerer}, {Schminovich}, {Shopbell}, {Taniguchi}, \& {Tyson}}]{Scoville2007}
{Scoville}, N., {Aussel}, H., {Brusa}, M., {et~al.} 2007, \apjs, 172, 1, \dodoi{10.1086/516585}

\bibitem[{{Smith} {et~al.}(2017){Smith}, {Croxall}, {Draine}, {De Looze}, {Sandstrom}, {Armus}, {Beir{\~a}o}, {Bolatto}, {Boquien}, {Brandl}, {Crocker}, {Dale}, {Galametz}, {Groves}, {Helou}, {Herrera-Camus}, {Hunt}, {Kennicutt}, {Walter}, \& {Wolfire}}]{Smith2017}
{Smith}, J.~D.~T., {Croxall}, K., {Draine}, B., {et~al.} 2017, \apj, 834, 5, \dodoi{10.3847/1538-4357/834/1/5}

\bibitem[{{Solanes} {et~al.}(2018){Solanes}, {Perea}, \& {Valent{\'\i}-Rojas}}]{Solanes2018}
{Solanes}, J.~M., {Perea}, J.~D., \& {Valent{\'\i}-Rojas}, G. 2018, \aap, 614, A66, \dodoi{10.1051/0004-6361/201832855}

\bibitem[{{Solomon} {et~al.}(1987){Solomon}, {Rivolo}, {Barrett}, \& {Yahil}}]{Solomon1987}
{Solomon}, P.~M., {Rivolo}, A.~R., {Barrett}, J., \& {Yahil}, A. 1987, \apj, 319, 730, \dodoi{10.1086/165493}

\bibitem[{{Stacey} {et~al.}(1991){Stacey}, {Geis}, {Genzel}, {Lugten}, {Poglitsch}, {Sternberg}, \& {Townes}}]{Stacey1991}
{Stacey}, G.~J., {Geis}, N., {Genzel}, R., {et~al.} 1991, \apj, 373, 423, \dodoi{10.1086/170062}

\bibitem[{{Stacey} {et~al.}(2010){Stacey}, {Hailey-Dunsheath}, {Ferkinhoff}, {Nikola}, {Parshley}, {Benford}, {Staguhn}, \& {Fiolet}}]{Stacey2010b}
{Stacey}, G.~J., {Hailey-Dunsheath}, S., {Ferkinhoff}, C., {et~al.} 2010, \apj, 724, 957, \dodoi{10.1088/0004-637X/724/2/957}

\bibitem[{{Stacey} {et~al.}(2018){Stacey}, {McKean}, {Robertson}, {Ivison}, {Isaak}, {Schleicher}, {van der Werf}, {Baan}, {Berciano Alba}, {Garrett}, \& {Loenen}}]{Stacey2018}
{Stacey}, H.~R., {McKean}, J.~P., {Robertson}, N.~C., {et~al.} 2018, \mnras, 476, 5075, \dodoi{10.1093/mnras/sty458}

\bibitem[{{Tazzari} {et~al.}(2018){Tazzari}, {Beaujean}, \& {Testi}}]{Tazzari2018}
{Tazzari}, M., {Beaujean}, F., \& {Testi}, L. 2018, \mnras, 476, 4527, \dodoi{10.1093/mnras/sty409}

\bibitem[{{Tielens} \& {Hollenbach}(1985)}]{Tielens1985a}
{Tielens}, A.~G.~G.~M., \& {Hollenbach}, D. 1985, \apj, 291, 722, \dodoi{10.1086/163111}

\bibitem[{{Umehata} {et~al.}(2017){Umehata}, {Matsuda}, {Tamura}, {Kohno}, {Smail}, {Ivison}, {Steidel}, {Chapman}, {Geach}, {Hayes}, {Nagao}, {Ao}, {Kawabe}, {Yun}, {Hatsukade}, {Kubo}, {Kato}, {Saito}, {Ikarashi}, {Nakanishi}, {Lee}, {Izumi}, {Mori}, \& {Ouchi}}]{Umehata2017}
{Umehata}, H., {Matsuda}, Y., {Tamura}, Y., {et~al.} 2017, \apj, 834, \dodoi{10.3847/2041-8213/834/2/L16}

\bibitem[{{Valentino} {et~al.}(2021){Valentino}, {Daddi}, {Puglisi}, {Magdis}, {Kokorev}, {Liu}, {Madden}, {G{\'o}mez-Guijarro}, {Lee}, {Cortzen}, {Circosta}, {Delvecchio}, {Mullaney}, {Gao}, {Gobat}, {Aravena}, {Jin}, {Fujimoto}, {Silverman}, \& {Dannerbauer}}]{Valentino2021}
{Valentino}, F., {Daddi}, E., {Puglisi}, A., {et~al.} 2021, \aap, 654, A165, \dodoi{10.1051/0004-6361/202141417}

\bibitem[{{van der Werf} {et~al.}(2010){van der Werf}, {Isaak}, {Meijerink}, {Spaans}, {Rykala}, {Fulton}, {Loenen}, {Walter}, {Wei{\ss}}, {Armus}, {Fischer}, {Israel}, {Harris}, {Veilleux}, {Henkel}, {Savini}, {Lord}, {Smith}, {Gonz{\'a}lez-Alfonso}, {Naylor}, {Aalto}, {Charmandaris}, {Dasyra}, {Evans}, {Gao}, {Greve}, {G{\"u}sten}, {Kramer}, {Mart{\'\i}n-Pintado}, {Mazzarella}, {Papadopoulos}, {Sanders}, {Spinoglio}, {Stacey}, {Vlahakis}, {Wiedner}, \& {Xilouris}}]{vanderWerf2010}
{van der Werf}, P.~P., {Isaak}, K.~G., {Meijerink}, R., {et~al.} 2010, \aap, 518, L42, \dodoi{10.1051/0004-6361/201014682}

\bibitem[{{Whitaker} {et~al.}(2017){Whitaker}, {Pope}, {Cybulski}, {Casey}, {Popping}, \& {Yun}}]{Whitaker2017}
{Whitaker}, K.~E., {Pope}, A., {Cybulski}, R., {et~al.} 2017, \apj, 850, 208, \dodoi{10.3847/1538-4357/aa94ce}

\bibitem[{{Wilner} \& {Welch}(1994)}]{Wilner1994}
{Wilner}, D.~J., \& {Welch}, W.~J. 1994, \apj, 427, 898, \dodoi{10.1086/174195}

\bibitem[{{Witstok} {et~al.}(2022){Witstok}, {Smit}, {Maiolino}, {Kumari}, {Aravena}, {Boogaard}, {Bouwens}, {Carniani}, {Hodge}, {Jones}, {Stefanon}, {van der Werf}, \& {Schouws}}]{Witstok2022}
{Witstok}, J., {Smit}, R., {Maiolino}, R., {et~al.} 2022, \mnras, 515, 1751, \dodoi{10.1093/mnras/stac1905}

\bibitem[{{Wolfire} {et~al.}(1990){Wolfire}, {Tielens}, \& {Hollenbach}}]{Wolfire1990}
{Wolfire}, M.~G., {Tielens}, A.~G.~G.~M., \& {Hollenbach}, D. 1990, \apj, 358, 116, \dodoi{10.1086/168966}

\bibitem[{{Zanella} {et~al.}(2018){Zanella}, {Daddi}, {Magdis}, {Diaz Santos}, {Cormier}, {Liu}, {Cibinel}, {Gobat}, {Dickinson}, {Sargent}, {Popping}, {Madden}, {Bethermin}, {Hughes}, {Valentino}, {Rujopakarn}, {Pannella}, {Bournaud}, {Walter}, {Wang}, {Elbaz}, \& {Coogan}}]{Zanella2018}
{Zanella}, A., {Daddi}, E., {Magdis}, G., {et~al.} 2018, \mnras, 481, 1976, \dodoi{10.1093/mnras/sty2394}

\bibitem[{{Zavala} {et~al.}(2021){Zavala}, {Casey}, {Manning}, {Aravena}, {Bethermin}, {Caputi}, {Clements}, {Cunha}, {Drew}, {Finkelstein}, {Fujimoto}, {Hayward}, {Hodge}, {Kartaltepe}, {Knudsen}, {Koekemoer}, {Long}, {Magdis}, {Man}, {Popping}, {Sanders}, {Scoville}, {Sheth}, {Staguhn}, {Toft}, {Treister}, {Vieira}, \& {Yun}}]{Zavala2021}
{Zavala}, J.~A., {Casey}, C.~M., {Manning}, S.~M., {et~al.} 2021, \apj, 909, 165, \dodoi{10.3847/1538-4357/abdb27}

\bibitem[{{Zhang} {et~al.}(2018){Zhang}, {Ivison}, {George}, {Zhao}, {Dunne}, {Herrera-Camus}, {Lewis}, {Liu}, {Naylor}, {Oteo}, {Riechers}, {Smail}, {Yang}, {Eales}, {Hopwood}, {Maddox}, {Omont}, \& {van der Werf}}]{Zhang2018}
{Zhang}, Z.-Y., {Ivison}, R.~J., {George}, R.~D., {et~al.} 2018, \mnras, 481, 59, \dodoi{10.1093/mnras/sty2082}

\bibitem[{{Zhao} {et~al.}(2016){Zhao}, {Yan}, \& {Tsai}}]{Zhao2016}
{Zhao}, Y., {Yan}, L., \& {Tsai}, C.-W. 2016, \apj, 824, 146, \dodoi{10.3847/0004-637X/824/2/146}

\end{thebibliography}
\clearpage
\appendix
\stepcounter{myequation}
\section{Disk Modeling Technical Addendum}
\label{sec:disktech}
\subsection{Data preparation}
\label{sec:dataprep}
\shorthandon
It is imperative that the data weights for the observed visibilities are set correctly. To accomplish that goal, we used the following data routines. First, \verb|initweights| is used with the parameter \verb|dowtsp=True| in order to generate weights for each spectral channel. Then the *CASA measurement set can be exported to *UVFITS. \footnote{If \texttt{WEIGHT\_SPECTRUM} in the *MS was unset, the weights in
the *UVFITS would be equal to the weights in the MS divided by the number of channels as explained in the *Casa Docs,
\url{https://casadocs.readthedocs.io/en/stable/api/tt/casatasks.data.exportuvfits.html},
and therefore too small by a factor of $n_{chan}$.
}

\subsection{Modeling and Results}

The nested sampling routine PyMultiNest \citep{Buchner2014,Feroz2009} was used to explore the parameter space of the model.
In each iteration of the nested sampling, KinMS \citep{Davis2013} is used to create a model rotating disk spectral cube in the image plane based on ten parameters. The image plane datacube is converted to visibilities using Galario \citep{Tazzari2018}, using the same baselines as the real \ALMA observations to ensure that the comparison between the model and data is as statistically robust as possible.

The full corner plot for our disk model is shown in \autoref{fig:corner}. Note that inclination angle and maximum velocity are strongly covariant, as expected from geometry.

\begin{figure}[p]
\centering
\includegraphics[width=\textwidth]{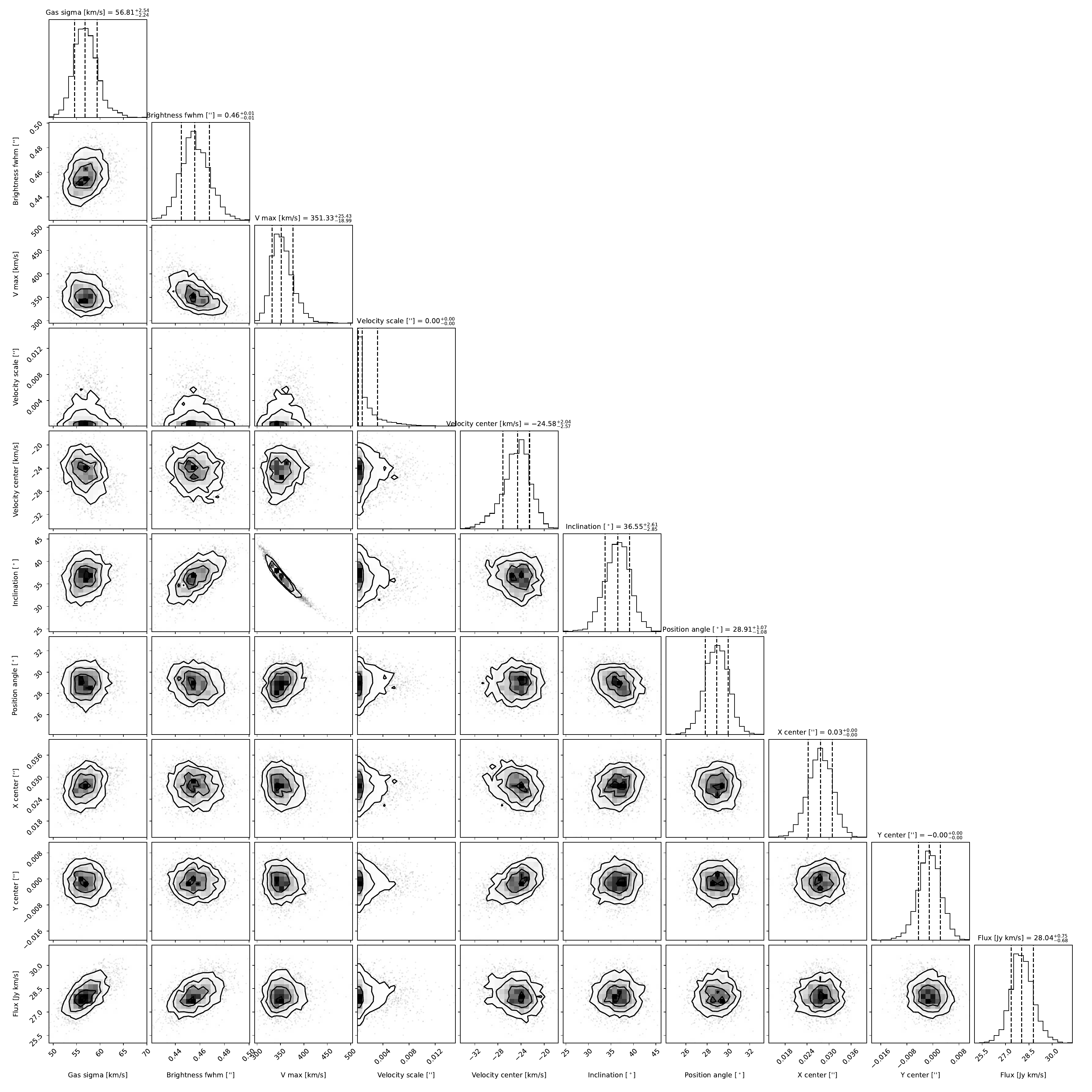}
\caption{Results from the UVModelDisk procedure. Here \sdssj is modeled as a rotating disk. Each panel is a histogram of the model parameters visited by the nested sampling algorithm, which is proportional to the posterior probability of those parameters. The 2-D histograms show the covariance of different parameters, while the 1-D histograms show the total probability distribution of that parameter. All parameters are well constrained, with the exception of the expected degeneracy between maximum circular velocity and disk inclination angle. Note also that, for the modeling procedure, the velocity zero point is set to the center velocity of the sub-cube used for modeling, rather than the prior redshift estimate, and the velocity sign convention is reversed due to channel ordering. This offset is $54$ km s$^{-1}$, which, when combined with the best model center velocity of $-25$ km s$^{-1}$, yields a center velocity estimate of $+80$ km s$^{-1}$ }
\label{fig:corner}
\end{figure}

\subsection{Residuals}
\label{sec:residuals}
In \autoref{fig:modelimage} we show an integrated image of the UVModelDisk model and also the residuals. Residuals were calculated by subtracting the model visibilities from the data visibilities and creating a dirty image of the result with \verb\tclean\. Note that the hole in the residual is not an indication that the flux was over-predicted, because it is surrounded by an annulus of positive flux. Instead it is merely an indication that the source shape is slightly flatter than Gaussian. During modeling we only considered emission from the main source, not the satellite, so the satellite is visible in the residual image (\autoref{fig:modelimage}, right).
\begin{figure}
\centering
\includegraphics[width=0.4\textwidth]{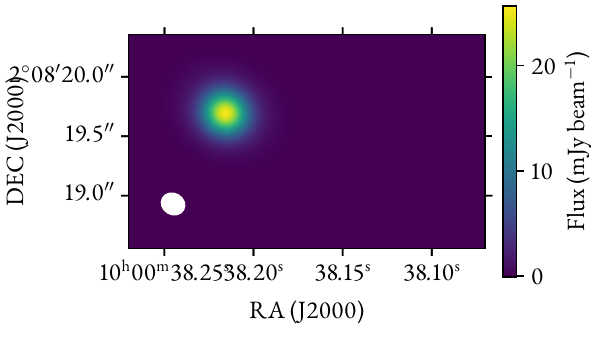}
\includegraphics[width=0.4\textwidth]{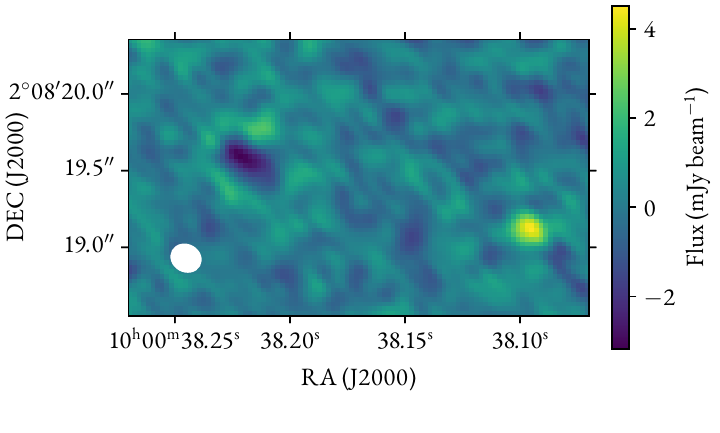}
\caption{[Left] Image of the best model from the UVModelDisk procedure. [Right] Dirty image of the residual visibilities from the best-fit model. This is an image of the (data visibilities - model visibilities). Both of these images were produced using the \texttt{tclean} task in *CASA on visibilities exported from UVModelDisk.
}
\label{fig:modelimage}
\shorthandoff
\end{figure}

\section{Modeling Cloud Size and Emission Filling Factors}
\label{sec:cloudsize}
We describe below how the observed \PDR and CO lines, and the far-\acro{IR} continuum flux and their ratios are used to derive cloud parameters within the \PDR paradigm including cloud size, numbers of clouds and the volume and area filling factors within a toy model that assumes spherical clouds externally heated by \FUV flux.

We begin by deriving the beam filling factor $\phi_{A}$. The ratios of relevant line fluxes are used within the PDR Toolbox to divide out beam filling factors. The raw output of the model line intensities assumes the source fills the beam, so that the ratio of the observed line intensity to the model outputs is the beam filling factor.  For a source at redshift $z$, the beam filling factor for the \cii line is therefore: $\phi_{A} = (1+z)^{4} (F_{\tx{\cii}}/\Omega(\tx{\cii}))/I_{\tx{\cii}}$. Here $F_{\tx{\cii}}/\Omega(\tx{\cii})$  is the observed intensity of the \cii line in a beam and  $I_{\tx{\cii}}$ is the line intensity given in the \PDRT models.  The factor of  $(1+z)^{4}$ comes from cosmological surface brightness dimming.  For an optically thick line emerging from a spherical cloud, the emitting surface area is $A_{cloud}=\pi r^{2}$. The single surface emission would also be true for the optically thin \cii line if the cloud were illuminated by a central source, so only one surface emits \cii line radiation.  However, if the clouds are spherical and uniformly illuminated externally by a pervasive radiation field, as one might expect for an extended star formation region, the surface area of each cloud for the optically thin line will be four times larger: $A_{cloud} = 4\pi r^{2}$. Using the observed \cii line flux, derived \cii line intensity from the \PDRT models and the beam size from \autoref{tab:toyparams} we find the beam area filling factor is:  $\phi_{A} = 4.2$.  There are typically 4 sheets of \cii emitting \PDR gas along each line of sight.

One may also derive the overall structure of the interstellar medium through a simple model of the \ISM based on the assumption that the CO line emission arises from molecular clouds and traces the molecular cloud mass, and the preponderance of the \cii line emission arises from \PDRs on the surfaces of these molecular clouds \citep{Wolfire1990}.  Here we make a simple model that assumes the \PDR gas number density is the same as that of the molecular cloud interior ($n_{H} \sim n_{H_2 })$, we have spherical clouds, all of which have the same molecular cloud radius, $r$, with a \cii emitting \PDR surface of depth $\Delta r$, and they are illuminated by a uniform \FUV field with strength $G$.  In such a model, the molecular cloud mass of each cloud is: $m_{m} = \frac{4}{3} \pi r^{3}\cdot n_{H{_2}} m_{H_{2}}$ and the mass of the \cii emitting surface layer is: $m_{PDR} = 4\pi r^{2} \Delta r\cdot n_{H} m_{H}$, so that the molecular to \PDR gas mass ratio is:

\begin{equation}
\label{eq:pdrgeometry}
\frac{m_m}{m_{PDR}}=\frac{\frac{4}{3}\pi r^3\cdot n_{H_2}\cdot m_{H_2}}{4\pi r^2\Delta r \cdot n_H\cdot m_H}=\frac{2r}{3\Delta r}
\end{equation}

If we assume all the clouds are identical and all the molecular and atomic gas is in the clouds, the ratio of total molecular gas to total atomic mass in the galaxy $M_m/M_{PDR} = m_m/m_{PDR}$. The column density of the \cii emitting region in a \PDR is related to both the \PDR density and incident \FUV field strength. For $n_{H}=6\times10^{3}$ cm$^{-3}$ and $G = 5 \times 10^{3}\; G_{0}$ the column density (assuming solar abundances) is $N_{H}= 4\times 10^{21}$ cm$^{-2}$ \citep{Wolfire1990}, so that $\Delta r\sim 4.0\times 10^{21} $ cm$^{-2}/n_{H} =0.22$ pc, and:
\begin{equation}
\label{eq:pdrscaling}
\frac{
M_{m}\left(\alpha_{CO}/0.8\right)
}{
M_{PDR}
}
=
\frac{ r\cdot n_{H}}{6.0\times 10^{21} \tx{ cm}^{-2}} =\left(\frac{1}{1940}\right)\left(\frac{r}{\tx{ pc}}\right) \left(\frac{n_{H}}{\tx{ cm}^{-3}}\right)
\end{equation}

To highlight the dependence of the following parameters on $\alpha_{CO}$, note that the molecular gas mass $M_m \propto \alpha_{CO}$ and \cite{Aravena2008} used $\alpha_{CO}=0.8 \tx{ M}_\odot$~(K km s$^{-1}$ pc$^2$ )$^{-1}$ to derive $M_{m}= 4.5\times 10^{10}\tx{ M}_\odot$. We have included a unitless scaling factor of $1=\left(\alpha_{CO}/0.8\right)$ where $\alpha_{CO}$ is important.
From our \PDR modeling,
$n_{H}= 6 \times 10^{3}$ cm$^{-3}$, and using our \cii flux with Eq. 2 from \cite{Wolfire1990}
$M_{PDR} \sim 4.4\times 10^{9}\tx{ M}_\odot$, so that the mass ratio is:
$M_{m}/M_{PDR} = 10 \left(\alpha_{CO}/0.8\right)$. Therefore, the radius of the CO emitting core of the cloud is:
$r_{CO}=3.4$ pc $\left(\alpha_{CO}/0.8\right)$. The total cloud radius is:
$r_{cloud}=r_{CO}+\Delta r = 3.6$~pc (since $r_{CO} \gg \Delta r$, to good approximation $r_{cloud} \propto \alpha_{CO}$). The mass of a single cloud core is:
$m_m=4\pi/3 r_{CO}^{3}\cdot n_{H_{2}} \cdot m_{H_{2}}= 4.7\times 10^{4}\tx{ M}_\odot \left(\alpha_{CO}/0.8\right)^3$.  Including the mass in the \PDR we get a total neutral gas mass of
$5.2\times 10^{4}\tx{ M}_\odot$ per cloud (and since $m_m \gg m_{PDR}$, $m_{tot} \propto \alpha_{CO}^3$ to good approximation).

The total number of clouds is the ratio of the total mass to the mass in a cloud:  $N_{clouds}=(4.83\times10^{10})/(5.2\times10^{4})=9.5\times 10^{5}$.  The \cii emitting region has an apparent radius of $0\farcs456$, which corresponds to a physical size of 3.85 kpc at the redshift of \jiooo (z = 1.8275) so that assuming a sphere, the total volume of the emission region is $2.3\times10^{11}$~pc$^{3}$. Within this volume there are $9.5\times10^{5}$ clouds, each of which occupies 192~pc$^{3}$ so the volume filling factor is $\phi_{vol}= 0.077\%$.   These model parameters are in very good agreement with models for \PDR geometry in unresolved systems presented in \cite{Wolfire1990}.

\end{document}